\documentclass[aps,floatfix,preprint,preprintnumbers,nofootinbib,superscriptaddress,natbib]{revtex4}
\usepackage{siunitx}
\usepackage[titletoc]{appendix}
\usepackage{graphicx,float}
\usepackage{epsfig}		
\usepackage{dcolumn}
\usepackage{bm}
\usepackage{subfigure}
\usepackage{lipsum}
\usepackage{fancyhdr}
\usepackage[all]{xy}
\usepackage{amsmath,upgreek}
\usepackage{amssymb}
\usepackage{empheq}
\usepackage{xcolor}
\usepackage{bbold}
\definecolor{lightgreen}{HTML}{90EE90}

\usepackage{pdfpages}
\usepackage{color}
\usepackage{graphicx,epstopdf}
\usepackage[colorlinks=true,
 linkcolor=red,
 urlcolor=purple,
 citecolor=blue,hyperindex]{hyperref}
\usepackage{cleveref}
\def\0{\mbox{\tiny $0$}}
\def\1{\mbox{\tiny $1$}}
\def\2{\mbox{\tiny $2$}}
\def\3{\mbox{\tiny $3$}}
\def\4{\mbox{\tiny $4$}}
\def\5{\mbox{\tiny $5$}}
\def\6{\mbox{\tiny $6$}}
\def\7{\mbox{\tiny $7$}}
\def\8{\mbox{\tiny $8$}}
\def\9{\mbox{\tiny $9$}}

\def\f14{\mbox{\tiny $\frac{1}{4}$}}

\def\L{\mbox{\tiny $L$}}

\def\R{\mbox{\tiny $R$}}

\def\mi{\mbox{\tiny $-$}}

\def\mt#1{\mbox{\textsl{#1}}}
\def\mbf#1{\mbox{\boldmath$#1$}}
\def\bb#1{\mbox{\footnotesize $(#1)$}}

\begin{document}
	\title{Revival patterns for Dirac cat states in a constant magnetic field}
\author{Caio Fernando e Silva}
\email{caiofernandosilva@df.ufscar.br}
\author{Alex E. Bernardini}
\email{alexeb@ufscar.br}
\affiliation{~Departamento de F\'{\i}sica, Universidade Federal de S\~ao Carlos, PO Box 676, 13565-905, S\~ao Carlos, SP, Brasil.}
	
	\date{\today}

\begin{abstract}
Considering the parity symmetry related to the Dirac equation, the interplay between energy localization and the temporal evolution of parity-defined quantum superpositions is investigated for fermions in a magnetic field. The unitary evolution of Dirac cat states is obtained by initializing either even or odd principal quantum numbers in the equivalent harmonic oscillator {\bf basis in relativistic Landau levels\bf}. Quantum operators feature well-defined selection rules for states thus identified, exhibiting a permanent revival structure. Our analysis is specialized for the survival probability function and for the expectation values of spinor matrix operators, which are identified as quantifiers of spin-parity correlations encoded in Dirac bispinors. In such a context, the time evolving quantum state also imprints a signature on the energy expansion. Namely, frequencies associated with revivals are doubled for each revival order, being observed up to a so-called {\em super revival} time scale. Results show that Dirac cat states exhibit a fractional revival structure, which works as a probe of suppressions and regenerations of intrinsic correlations driven by the discrete spin-parity degrees of freedom of Dirac bispinors.
\end{abstract}
	
\maketitle 

\section{Introduction}

The quantum revival is a general feature of the long-time evolution of quantum systems, which describes a quantum state that regenerates sufficiently close to its initial configuration. Nonrelativistic and relativistic localized quantum wave packets exhibit revivals due to quantum interference and have been studied in atoms, molecules, and Bose-Einstein condensates, both theoretically \cite{Banacloche,Robert1, Robert2,Strange,Robinett,Romera,Mather} and experimentally \cite{JinShi,Robinett,Greiner2,Rempe}. In particular, quantum revivals can be implemented by structures that mimic the Dirac equation \cite{Gerritsma,Shore, Bermudez, Krueckl}, evincing the connection of such Dirac-like intrinsic structures with the evolution of wave packets as well as with physically measurable phenomena that include revivals of electron currents and phase transitions in such systems \cite{Romera2,Zondiner,Romera3}. 

Considering that the Dirac equation supports a $SU(2)\otimes SU(2)$ group structure related to the spin-parity degrees of freedom \cite{MeuPRA}, Dirac bispinors can be regarded as two-qubit entangled structures \cite{extfields,n010}, eventually correlated by phase-space variables. In a preliminary context, the algebraic properties of the Dirac-like Hamiltonians have already been investigated for generalized Poincar{\'e} classes of external constant potentials \cite{Thaller,extfields} required for mapping Dirac bispinors onto controllable physical systems \cite{
n001, n002, n004, n005, n006, graph03, graph04, CastroNeto,MeuPRB,PRBPRB18,New} and high energy physics scenarios \cite{n010,Bernardini:2011fph,EPJC21}. Since the elementary information profile of bispinors is modified by localization effects due to potentials that break the symmetries implied by the Dirac equation \cite{BernardiniEPJP,Greiner}, the inclusion of continuous degrees of freedom \cite{Bermudez,Rusin} has also been considered in the description of quantum correlations in Dirac-like systems. 
Of course, localization features are also encompassed by the elementary information content and quantum correlations involving generic discretized qubit systems, which have long been noticed as an essential structure for developing quantum correlation tasks \cite{Braunstein,Steane,Ekert}. 
In addition, the thorough control of quantumness and quantum correlation quantifiers \cite{Auyuanet, Vedral,Henderson,n024,Silva}, in such a context, may be engendered if one intends to implement them in more realistic settings where a quantum system is coupled to an external environment \cite{Braunstein,Mintert}. 

In fact, although environmental interactions in open systems are the main source of dissipation of quantum correlations \cite{Zurek}, even closed systems can exhibit similar properties, for instance, quantified by some kind of entanglement sudden death \cite{Ting, Cui}, i.e. a particular entanglement suppression that occurs in a finite time and is strongly dependent on initial conditions.
For addressing the possibility of entanglement sudden death and revivals for Dirac-like systems driven by continuous degrees of freedom, the quantum correlation properties are mostly encompassed by the computation of quantum observables using the Weyl-Wigner framework \cite{Weickgenannt, Zhuang,1986,1987, Wigner, Case}. This has opened up an interesting path for computing elementary correlations in Dirac-like localized systems, which, for instance, emerge with the introduction of localization constraints imposed by constant magnetic field couplings. In particular, in such a scenario, which includes mesoscopic coherent superposition in relativistic Landau levels \cite{Bermudez}, Dirac cat states and their associated quantum correlation information have already been investigated \cite{PRA2021}. 

{\bf
Such preliminary results suggest that the entanglement between spinors and orbital wave functions could create intrinsic (intraparticle) spin-parity correlations similar to those introduced and discussed in Refs.~\cite{extfields,MeuPRA} for Dirac spinors driven by constant external potentials (see the Appendix \ref{AppA}). In the present case, different from Refs.~\cite{extfields,MeuPRA}, due to the presence of a position dependent localizing quantum potential, the associated Dirac equation solutions are cast as a tensor product between spin-parity (two discrete degrees of freedom) and orbital (one degree of freedom) quantum states.  Thus, even departing from null spin-parity quantum correlations, the spin-parity intrinsic entanglement between such discrete degrees of freedom can emerge from internal correlations with the continuous degree of freedom driven, for instance, from the temporal evolution of Gaussian localized Dirac cat states. 

Considering such aspects, the main proposal of this work is to identify time-persistent spin-parity intrinsic entanglement patterns in relativistic Landau levels at long time scales.} The emergence of these time scales in the quantum concurrence profile suggests the emergence of wave packet revivals as well as the identification of a semiclassical behavior related to states that are energy localized but are also in a superposition of arbitrarily distant wave packets. 
In fact, probing the dynamics of parity-defined states with the machinery developed for quantum revivals is expected to reveal additional classifications for the time scales \cite{Aronstein} which drive the phenomena. 
Therefore, our final aim is to identify the quantum observables, namely those related with the energy spectral function and the survival probability, which quantitatively detect the characteristic revival structure of Dirac cat states. The conditions on the energy parameters to detect and classify revival time scales shall be investigated in terms of the variables that control the weak- and high-field limits, driving the correlation profile of Dirac spinors. The correspondence with the temporal evolution of spinor matrix operators will be necessarily investigated. The collapse and revival pattern is then expected to be related to disentanglement and regeneration of quantum correlations implied by Dirac cat states here considered.

The paper is then organized as follows. In section II, Dirac cat states are obtained from spin and parity considerations, and their energy spectral function is described. In section III, the corresponding survival probability is obtained from the eigenstate expansion and it is shown to exhibit several revival orders under wave packet evolution. In section IV, revivals are also identified for quantum correlation observables obtained from the corresponding Weyl-Wigner phase-space framework. In particular, the oscillation between entangled and disentangled states is shown to be related to collapse and revival patterns. Finally, in section V, the results are summarized, and possible extensions pointing to low-dimensional system phenomenology is suggested.

\section{Energy expansion of Dirac cat states in constant magnetic fields }

The Hamiltonian for a charged fermion coupled with a constant magnetic field can be written as 
\begin{equation}
H = \mbox{\boldmath$\alpha$} \cdot ({\bf p} + (-1)^r\, e {\bf A}) + \gamma_0 M,
\label{eq0}
\end{equation}
where one has assumed natural units, with $c=\hbar=1$, $e$ as the positive unit of charge, and that the potential vector, ${\bf A}$, results into the magnetic field ${\bf B} = \mbox{\boldmath$\nabla$} \times {\bf A}$, with $r = 1$ and $2$ labeling the positive and negative intrinsic parity states\footnote{{\bf
See the Appendix \ref{AppA} for an explanation of the meaning of the intrinsic parity for Dirac spinors. }}, respectively. For the gauge chosen as ${\bf A} = \mathcal{B}\,x\, \hat{\bf y}$, which corresponds to a magnetic field along the $z$-direction, a set of orthogonal Dirac Hamiltonian eigenstates from \eqref{eq0} can be written as \cite{Bagrov}\footnote{
{\bf Despite the simplifying notation, the signs coming from $(-1)^r$ in Eqs.~\eqref{eq0} and \eqref{stationarysolutions} have
different physical meanings. In fact, one can have different charges with the
same energy sign. Here, the solutions correspond to a particular gauge driven by $\vec A$ for a
uniform magnetic field along the z-axis. Given that $r = 1,\,2$ is also related to even/odd parity solutions, respectively,  it is relevant to observe that for a gauge which gives rise to circular symmetric solutions, such notation shall be ruined.}}
\begin{equation}
\psi = \exp\big[i((-1)^r E_n t + k_y y + k_z z)\big] u_{n,r} ^\pm (s_r), \label{stationarysolutions}
\end{equation}
i.e. plane-wave solutions in both $y$ and $z$ directions, with the energy eigenvalues corresponding to the $n$-th Landau level identified by
\begin{equation}\label{eigenvalue}
 E_n =  \sqrt{M^2 + k_z ^2 + 2n e \mathcal{B}}.
\end{equation}
To summarize the influence of the magnetic field, the dynamics along the $x$-coordinate is shifted according to
\begin{equation}
s_r = \sqrt{e {\mathcal B}} \left( x + (-1)^r\frac{ k_y }{ e{\mathcal B}}
\right),
\label{222}
\end{equation}
such that the positive parity ($r=1$) space-dependent spinors can be written as
\begin{eqnarray}\label{9998}
u^+_{n,1}(s_1) = \sqrt{\eta_{n}}\left( \begin{array}{c} 
\mathcal{F}_{n-1}(s_1) \\ 0 \\ 
A_{n}\, \mathcal{F}_{n-1}(s_1) \\
-B_{n}\, \mathcal{F}_{n} (s_1) 
\end{array} \right), \quad 
u^-_{n,1}(s_1) = \sqrt{\eta_{n}}\left( \begin{array}{c} 
0 \\ \mathcal{F}_{n} (s_1) \\
-B_{n}\,
\mathcal{F}_{n-1}(s_1) \\ 
-A_{n}\,\mathcal{F}_{n}(s_1)
\end{array} \right), \quad 
\end{eqnarray}
and the negative parity ($r=2$) ones as
\begin{eqnarray}
u^+_{n,2}(s_2) = \sqrt{\eta_{n}}\left( \begin{array}{c} 
B_{n}\,
\mathcal{F}_{n-1}(s_2) \\ 
A_{n}\,\mathcal{F}_{n}(s_2) \\ 
0 \\ \mathcal{F}_{n} (s_2)
\end{array} \right), \qquad 
u^-_{n,2}(s_2)= \sqrt{\eta_{n}}\left( \begin{array}{c} 
-A_{n}\,
\mathcal{F}_{n-1}(s_2) \\ 
B_{n}\,
\mathcal{F}_{n} (s_2) \\ 
\mathcal{F}_{n-1}(s_2) \\ 0
\end{array} \right),
\label{9999}
\end{eqnarray}
where the dimensionless parameters $A_n$, $B_n$, and $\eta_n$ are given by \cite{BernardiniEPJP,PRA2021}\footnote{One notices that the set of parameters above were introduced in dimensionless form so as to simplify the analysis of typical regimes that dictate the influence of the rest mass and kinetic energy as compared to the magnetic field intensity. For instance, in the rest frame of the particle with $A_n = 0$, weak magnetic fields are described by $B_n \approx 0$, if the energy is dominated by the rest mass. This establishes a connection with accessible experiments that allow a fine control over these parameters, which would be otherwise inaccessible, as in the case for high magnetic fields $B_n \approx 1$ for massive fermions.}
\begin{equation} \label{parameters}
A_n = \frac{k_z}{E_n + M}, \quad B_n = \frac{\sqrt{2n\, e \mathcal{B}}}{E_n + M}, \quad \eta_n = \frac{E_n + M}{2E_n},
\end{equation}
satisfying the constraints $0 \leq A_n$, $B_n \leq 1$ and $\eta_n(A_n ^2 + B_n ^2 + 1) =1$,
with the functions $\mathcal{F}_n (s_r)$ describing normalized Hermite polynomials, $H_n (s_r)$, as
\begin{equation}
\mathcal{F}_n (s_r) = \left( \frac{\sqrt{e \mathcal{B}}}{n! \, 2^n \sqrt{\pi}}\right)^{1/2} e^{-(s_r) ^2/2} H_n (s_r),
\end{equation}
which form an orthonormal basis \cite{Gradshteyn}, where the principal quantum numbers $n$ label the harmonic oscillator (HO) basis. 

{\bf To define states with special symmetries in the magnetic field, one notices that the Hamiltonian eigenstates are also eigenstates of the total parity operator $\hat{P}$ \cite{Greiner} for $k_y = k_z = 0$, which acts on both, spatial and spinor, components of a wave function as }
\begin{equation}
 \hat{P} \phi (\mathbf{x},t) = \gamma_ 0 \phi (-\mathbf{x},t).
\end{equation}
Since Hermite polynomials have parity symmetry, the basis for a constant magnetic field consists of states that, separately, are neither eigenstates of the intrinsic parity operator nor eigenstates of the continuous parity operator. This is the parity invariance that follows from the Dirac equation. However, particular spinor configurations have, at least initially, well-defined intrinsic parity and continuous parity. Additional symmetries can excite eigenstates that determine the temporal evolution of the quantum superposition. Dirac cat states (as reported in Appendix \ref{AppB} \cite{PRA2021}) have an interesting behavior in that they select alternating eigenstates, that is, principal quantum numbers $n$ that have the same parity. {\bf  Explicitly, these states are given by
\begin{equation}\label{catstates}
\phi^{S,A} (s, t=0) = \frac{1}{2}\left( \frac{e \mathcal{B}}{\pi} \right)^{1/4} \bigg \{ \exp\left[-\frac{1}{2}(s-a)^2 \right] \pm \exp\left[-\frac{1}{2}(s+a)^2 \right] \bigg \} \left(\begin{matrix} 1 & 0 & 0& 0 \end{matrix}\right)^T,
\end{equation}
where $S(A)$ stands for a symmetric (antisymmetric) cat state and corresponds to the ``+"(``-") sign. In brief, cat states exhibit particular symmetries that will be investigated now by studying their energy content. }
\subsection{Energy distribution of cat states}

\begin{figure} [H]
 \centering
 \includegraphics[width=1\textwidth]{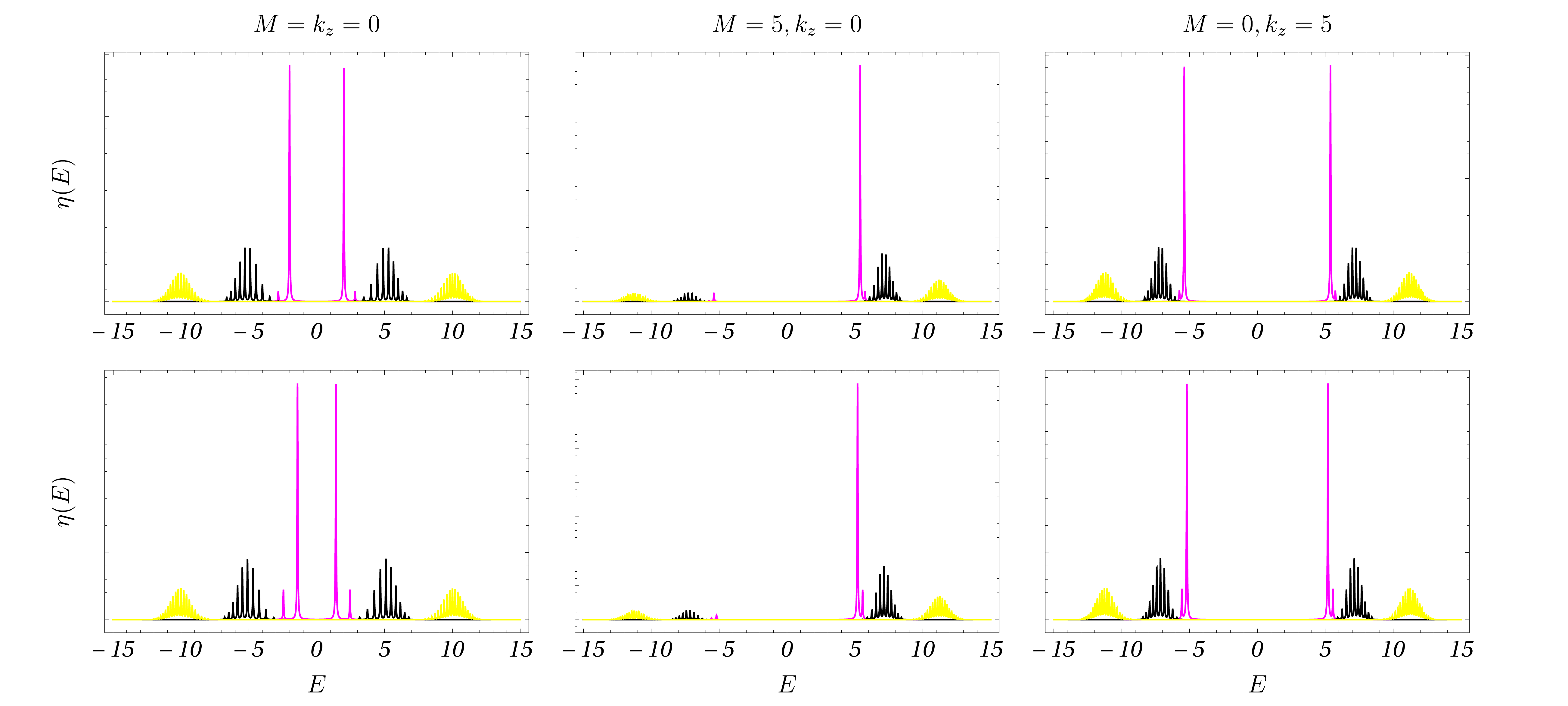} 
 \\
 \caption{Energy distribution of antisymmetric (top) and symmetric (bottom) cat states (cf. Eqs.~(\ref{levela0})-(\ref{levela2})) dependent on the distance parameter $a=1$ (magenta, gray), $5$ (black), $10$ (yellow, light gray) {\bf  around a mean energy level}. The $y$-axis is unscaled, and plotted quantities are dimensionless (cf. (\ref{eigenvalue})). }
 \label{powerspectrum}
\end{figure}

The parity of these states is noticed in the overlap between the initial state and the Hamiltonian eigenstates, which is described by the spectral function \cite{Ballentine}, 
\begin{equation}\label{spectral}
 \eta (E) = \langle \phi ^{A,S} \vert \delta (E - H) \vert \phi ^{A,S} \rangle = \sum _{\nu,r,m} \vert c_{m,r} ^\nu \vert ^2 \delta( E - (-1)^r E_m),
\end{equation}
for either symmetric or antisymmetric $\vert \phi ^{A,S} \rangle $ Dirac cat states separated by a distance parameter $a$. It yields the Hamiltonian eigenvalues that dictate the state time evolution.{\bf Thus, it provides the energy distribution of a particular quantum state. Whereas for a stationary solution the spectral function is a trivial delta function, time-dependent superpositions have a finite width energy distribution. In fact, if the spectral function is highly localized around a large quantum number, the quantum state oscillates with well-defined time scales \cite{Romera2}. Fig.~\eqref{powerspectrum} depicts that antisymmetric and symmetric cat states have a similar spectral function, except for states close to the origin as $a\rightarrow 0$.  Moreover, the spectral function is centered around an average energy value, from which one can identify a mean Landau level from the eigenvalue expression in Eq.~\eqref{eigenvalue}.

The Hamiltonian parameters control the shape of the energy distribution, which is plotted for $e\mathcal{B} = 1$. One notices that an increasing mass selects energy eigenvalues of the same sign, suppressing the interference of opposite energies. The plot for $M=5$ shows that only eigenstates with $E>0$ contribute to this particular state.  Another important parameter to be studied is the distance $a$, which shifts the center of the energy distribution, i.e., the mean Landau level, of the quantum superposition. As $a\rightarrow \infty$, symmetric and antisymmetric cat states are distinguishable only by an initial phase factor. In either case, HO quantum numbers have the same parity. }
Besides the alternating excited Landau levels, cat states exhibit an energy localization that depends on the phase-space distance parameter. As more eigenstates contribute to the quantum superposition, it spreads along a larger energy interval. In this case, the distribution of the coefficients squared for large quantum numbers is fitted to a Gaussian distribution
\begin{equation} \label{gaussiandist.}
 P_n = \frac{1}{\Delta n \pi^{1/2}} \exp\left[-\frac{1}{2}\left(\frac{n-n_0}{\Delta n/\sqrt{2}}\right)^2 \right],
\end{equation}
for each energy sign, and the normalization follows from the parity of $n$. The spread $\Delta n$ also depends on the distance parameter and determines the accuracy of expanding energy eigenvalues around the mean value. If $M=k_z=0$, the Landau level spacing is more evinced and tends to $\sqrt{2n} - \sqrt{2(n-2)}$. Typical calculations for a well-defined long time behavior restricts $\Delta n$ to $\Delta n/ n_0 \ll 1$ \cite{Robinett, Krueckl, Romera}. Such restrictions will be discussed in the following, as these parameters can be simply fitted to the infinite expansion of the hyperbolic functions used in Eqs.~(\ref{levela0})-(\ref{levela2}). 

\subsection{Time evolution of energy-localized states}

The energy distribution, depicted by the spectral function above, and the dynamical evolution of the quantum system have a one-to-one correspondence that can be explored in order to elicit useful information from the symmetry-defined Dirac cat states. When considering dynamical variables, which exhibit a time-dependence of the form of $\exp[i(-1)^r E_n t]$ for each excited eigenstate, the overall collective motion does not seem at first to exhibit any relevant pattern. If the Hamiltonian eigenvalues are not too spread out about the mean energy, it is quite the opposite. {\bf By expanding the energy of any eigenvalue $ E_n \equiv E(n)$ in the cat state around a mean Landau level (cf. Fig.~(\ref{powerspectrum})), one has }
\begin{equation}\label{enexpansion}
E(n) \approx E(n_0) + E'(n_0)(n-n_0) + \frac{E''(n_0)}{2}(n-n_0)^2
+ \frac{E'''(n_0)}{6}(n-n_0)^3 + \cdots.
\end{equation}
Each order in the energy expansion defines a characteristic frequency, which in turn modulates the higher frequency oscillations.
It is worth emphasizing that the energy spread depends not only on the principal quantum number, $n$, but also on the mass and kinetic parameters, $M$ and $k_z$. Hence, if the energy distribution is squeezed around the mean value, the energy expansion is more accurate. 

The first frequency, $E_{n_0}$, is a constant phase for quantum states whose energy eigenvalues have a definite sign. For Dirac cat states in Fig.~\eqref{powerspectrum}, this corresponds to the case where the mass dominates. Otherwise, it is a typical interference for states with opposite intrinsic parity, which, in some sense, is analogous to the Zitterbewegung effect in relativistic quantum mechanics \cite{Itzykson} and is always regenerated in a quantizing magnetic field. This rapid oscillation changes the phase factor of cat states throughout their time evolution, which is not observed for Schr{\"o}dinger-like quantum revivals, since all energy (frequency) eigenvalues have the same sign. {\bf This is one of the distinguishing features of the time-evolution in Dirac systems.}

As other time scales are considered, both intrinsic parities contribute to the long time evolution. This follows from the energy spectral function and by noticing that higher order terms in Eq.~\eqref{enexpansion} determine the temporal evolution even if the negative and positive signs of energy eigenvalues are found in the quantum superposition. Thus, the previous oscillation pattern is suppressed temporarily. An interesting feature in the energy expansion of both symmetric and antisymmetric cat states is that each term $(n-n_0)$ has also defined parity, and time scales associated to each frequency will be re-scaled by powers of 2. For instance, the other three periods are given by 
\begin{equation}
T_{1} = \frac{2\pi \hbar}{2|E'(n_0)|}
\quad
,
\quad
T_{2} = \frac{2\pi \hbar}{4|E''(n_0)|/2}
\quad
, 
\quad
\mbox{and}
\quad
T_{3} = \frac{2\pi \hbar}{8|E'''(n_0)|/6}
\, .
\label{timescales}
\end{equation}
{\bf In the usual semiclassical approach, they originally correspond to the so-called {\em classical}, {\em revival}, and {\em super revival} time scales \cite{Robinett,Krueckl,Romera,Romera2}. However, one should notice that even if these revivals are similar to the patterns observed in localized wave packets, the redefinition of time scales is a consequence of the parity symmetry of the quantum superposition, which only excites half of the Landau levels, when compared to a standard localized quantum state. Thus, cat states regenerate faster with well-defined time scales. From the expression of the cat states in Eq.~\eqref{catstates}, a Gaussian state can be obtained simply by the superposition of symmetric and antisymmetric states. The resulting Gaussian state oscillates with periods given by $2T_1, 4T_2, 8T_3$, respectively, when comparing to Eq.~\eqref{timescales}. }

The classical period corresponds to the oscillation that envelopes the previous $E_{n_0}$ frequency, when present. At each period, this higher frequency is regenerated, and the spread of the wave packet increases. After a few oscillations, the quantum state enters in the collapsed state more or less rapidly depending on the number of excited eigenstates. Semiclassical equations of motion are typically derived from the wave packet center (the center-of-mass) and momentum, which justify the classical correspondence \cite{Robinett,Krueckl}. However, if eigenfunctions with both energy signs have a significant overlap, their interference needs to be taken into account and the semiclassical motion might not be obvious depending on the particular observable chosen. {\bf This has been discussed for the Dirac oscillator and low-dimensional Dirac systems in magnetic fields \cite{Romera,Romera2}, for which revivals have been described in electron currents. It has been observed that the regeneration of the quasiclassical current in graphene is never devoid of the Zitterbewegung described above, since there is a contribution of positive and negative Landau levels. Moreover, as revivals depend on the localization of the wave packet, the increasing broadening of Landau levels might destroy the regeneration of currents if the magnetic field is not strong enough.

To compare these results with a strictly non-classical behavior, the quantum superposition considered here always has both energy signs for non-negligible mass. In fact, the symmetry-defined states are not even localized when not centered at the origin (cf. Eqs.~(\ref{levela0})-(\ref{levela2})). Therefore, there is no proper identification with a semiclassical behavior for a quantum particle in a magnetic field.} In this context, the velocity operator will also be discussed in Sec. \ref{sec4}, exploring further the motion of cat states as the superposition of two position-localized states, which are also quantum superpositions. Nevertheless, the classical period can still be defined for cat states as a direct consequence of the energy localization, and one shall keep the notation $T_1$, half the original classical time scale, since there is no classical correspondence in the system evolution. 

The next scale is identified by $T_2$, the revival time, at which the initial state is reformed after oscillating with the initial periodicity $T_1$ and collapsing afterwards. Once again, the $T_1$ periodicity is observed. The time-dependence of the exponential at second order in the energy expansion (cf. Eq.~\eqref{enexpansion}) thus approximately returns to unity. Even more interestingly is the occurrence of fractional revivals, which are observed at fractions of $T_2$, when smaller correlated copies of the initial state relocalize and oscillate with the same periodicity. However, each copy is temporally shifted, and thus the overall observed temporal evolution has periodicity given by fractions of the initial period $T_1$. If the state is not exactly reformed at the revival scale, the super revival $T_3$ can be observed. It follows the same general pattern, where the state is partially reformed and exhibits all the previous oscillation periodicities, with typical fractions of the super revival time to be defined later.

The features identified above can all be observed in the time evolution of energy-localized quantum superpositions with a discrete spectrum. To clear up this property, the evaluation of the associated time-dependent survival probability will be discussed in the following.

\section{Survival probability}

The complementary roles of the energy localization and time evolution descriptions of quantum states shall be further explored. From the time-evolved cat states,
\begin{equation}
 \vert \phi ^{A,S} (s, t) \rangle = e^{-i H_0 t} \vert \phi ^{A,S} (s, t=0)\rangle,
\end{equation}
the absolute value of the survival amplitude that the system remains in the original state $\vert \phi ^{A,S} (s, t=0)\rangle$ is
\begin{equation}
\vert \mathcal{C}(t) \vert = \vert \langle \phi ^{A,S} (s, 0) \vert e^{-i H_0 t} \vert \phi ^{A,S} (s, 0) \rangle \vert,
\end{equation}
which can be regarded as an orthogonality measure for a cat state ensemble \cite{Hilgevoord}. The revival structure described previously can thus be detected through this function, yielding a non-zero amplitude each time a smaller copy of the initial state approaches the same point in phase space, that is, the state is partially regenerated. Then, the initial state can be expanded in the Hamiltonian eigenfunctions (cf. Eqs.~\eqref{levela0}-\eqref{levela2}), 
\begin{eqnarray} \label{autocorrelation}
 \mathcal{C}(t) &=& \sum _{\nu,r,n} \vert c_{n,r} ^\nu \vert ^2 e^{(-1)^r i E_n t} \nonumber \\
 &=& \sum _{\nu,r,n} \vert c_{n,r} ^\nu \vert ^2 \int \, dE\, e^{-iE t} \delta( E - (-1)^r E_n) \nonumber \\
 &=& \int \, dE\, e^{-iE t} \bigg( \sum _{\nu,r,n} \vert c_{n,r} ^\nu \vert ^2\delta( E - (-1)^r E_n) \bigg), 
\end{eqnarray}
where the expression inside the parentheses is the spectral function in Eq.~\eqref{spectral}, and the function $\mathcal{C}(t)$ is usually referred as the autocorrelation function in the framework of wave packet revivals \cite{Robinett, Krueckl}. The equivalence between the energy distribution and the corresponding time evolution is established by the Fourier transform, which extracts the excited eigenstates from the temporal evolution. Therefore, the long time oscillation pattern is more defined for a localized energy distribution. In physical applications, where the spectral function is not infinitely localized as a sum of Dirac delta functions due to additional interactions, $\mathcal{C}(t)$ also describes finite lifetimes of unstable excited states \cite{Ballentine}.

The normalization of the Gaussian-distributed expansion coefficients constrains it to $\vert \mathcal{C}(t) \vert \leq 1 $, with a time dependence simply given by the eigenfunction exponentials. Thus, the time scales introduced in Eq.~\eqref{timescales} describe exactly the long-time behavior of $\mathcal{C}(t)$, yielding peaks when the state is recovered. Turning now to the computation of the parity-defined revival structure for Dirac cat states, one evaluates the survival probability for four relevant time scales defined by the energy expansion, up to the super revival time scale. 

\subsection{Collapse and revivals}

\begin{figure}
 \centering
 \includegraphics[width=0.5\textwidth]{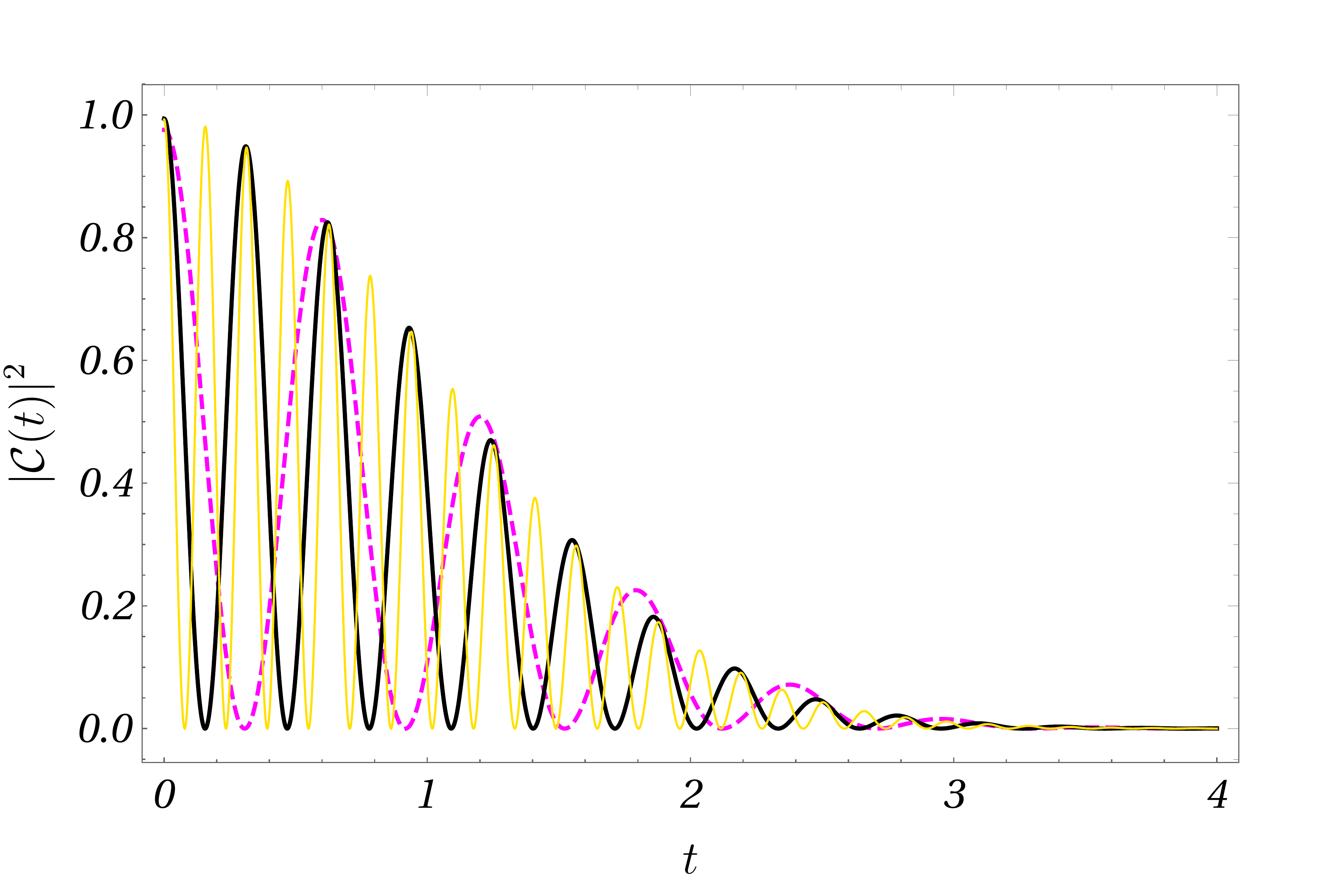}.
 \\
 \caption{Interference between states with positive and negative intrinsic parity at short time scales for increasing distance parameter: $a = 5,\, 10$ and $20$ for dashed magenta (gray), continuous black, and continuous yellow (light gray) lines, respectively, where the one-particle parameters (see Eq.~\eqref{parameters}) are fixed at $A_{n_0}= 0, B_{n_0}= 1$, and $\eta_{n_0} =1/2 $ (massless limit). }
 \label{autocorrelation1}
 \label{zbmodulation}
\end{figure}

The initial time development for several distance parameters $a$ is depicted in Fig.~\eqref{zbmodulation} in the rest frame of the particle. It is worth noticing that from now on the time parameter $t$ is also plotted as a dimensionless quantity (see Eq.~\eqref{autocorrelation}).  The ratio $M^2/e\mathcal{B} \approx 0$ simulates high-field limits for massive particles, which means that time scales in the quantum system evolution are shortened. As the HO number spreads in the cat state with increasing $a$, one clearly sees additional frequencies in the initial evolution. They decay in time but are observed again at the next time scale. In fact, the interference between positive and negative intrinsic parities is always regenerated when the quantum state is partially reformed. 

\begin{figure} [H]
 \centering
 \includegraphics[width=0.6\textwidth]{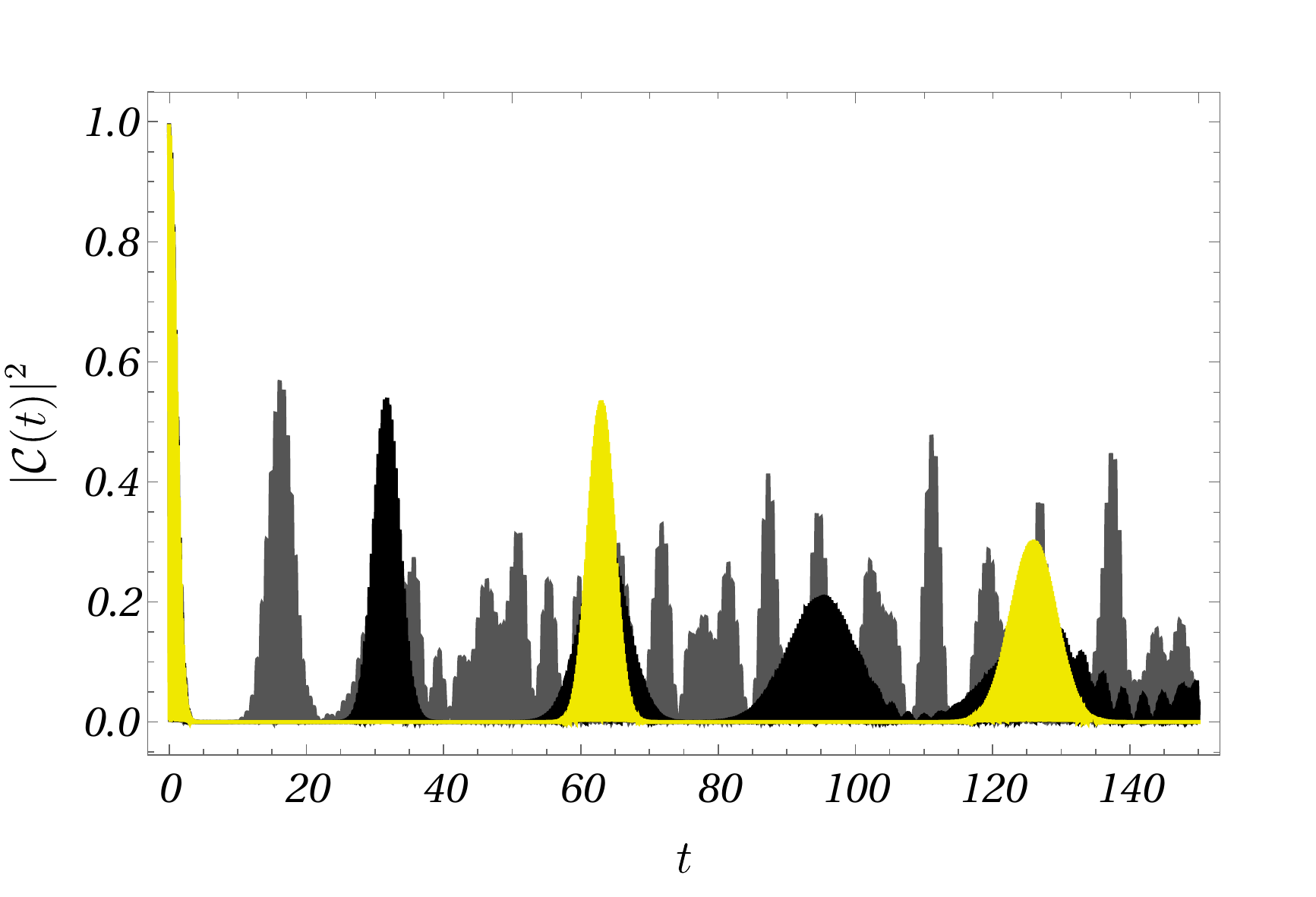}.
 \\
 \caption{Survival probability at the scale of $T_1$, with $a=5$ ($T_1 = 15$), 10 ($T_1 = 30$), 20 ($T_1 = 63$) for gray, black, and yellow (light gray) lines, respectively, where the one-particle parameters (see Eq.~\eqref{parameters}) are fixed at $A_{n_0}= 0, B_{n_0}= 1$, and $\eta_{n_0} =1/2 $. All plots overlap partially at $t=0$.}
 \label{t1scale}
\end{figure}

The initial behavior from Fig.~\eqref{zbmodulation} describes only the evolution at short times, since this rapid oscillation is enveloped by another oscillation with periodicity $T_1$. Once again, the amplitude decreases until the wave function spreads significantly, and the system no longer recovers the initial oscillation, now without any obvious frequency (cf. Fig.~\eqref{t1scale}). The decreasing amplitude resembles the exponential decay law found in Lorentzian spectral functions \cite{Bonitz}, since the eigenenergies are clumped together. However, after some time, the state regenerates partially and a local periodic motion can once again be observed at the scale of $T_2$. In particular, this is evident for $a=5$ cat states, for which $\mathcal{C}(t)$ only oscillates twice before collapsing. The collapse and regeneration will continue indefinitely, as it will be discussed for other time scales. The same pattern is also observed for greater values of the distance parameter $a$, as longer times are considered according to Eq.~\eqref{gaussiandist.}. It is emphasized that the scale $T_1$ is actually half the period estimated for the evolution of localized wave packets. The period associated to the corresponding classical oscillation is $T_{CL} \approx 30, 60, 126 $ for the average Landau level $n_0 = 12, 50, 200$, respectively. Therefore, cat states oscillate with half the classical periodicity, reforming the initial state.

\begin{figure}
 \centering
 \includegraphics[width=\textwidth]{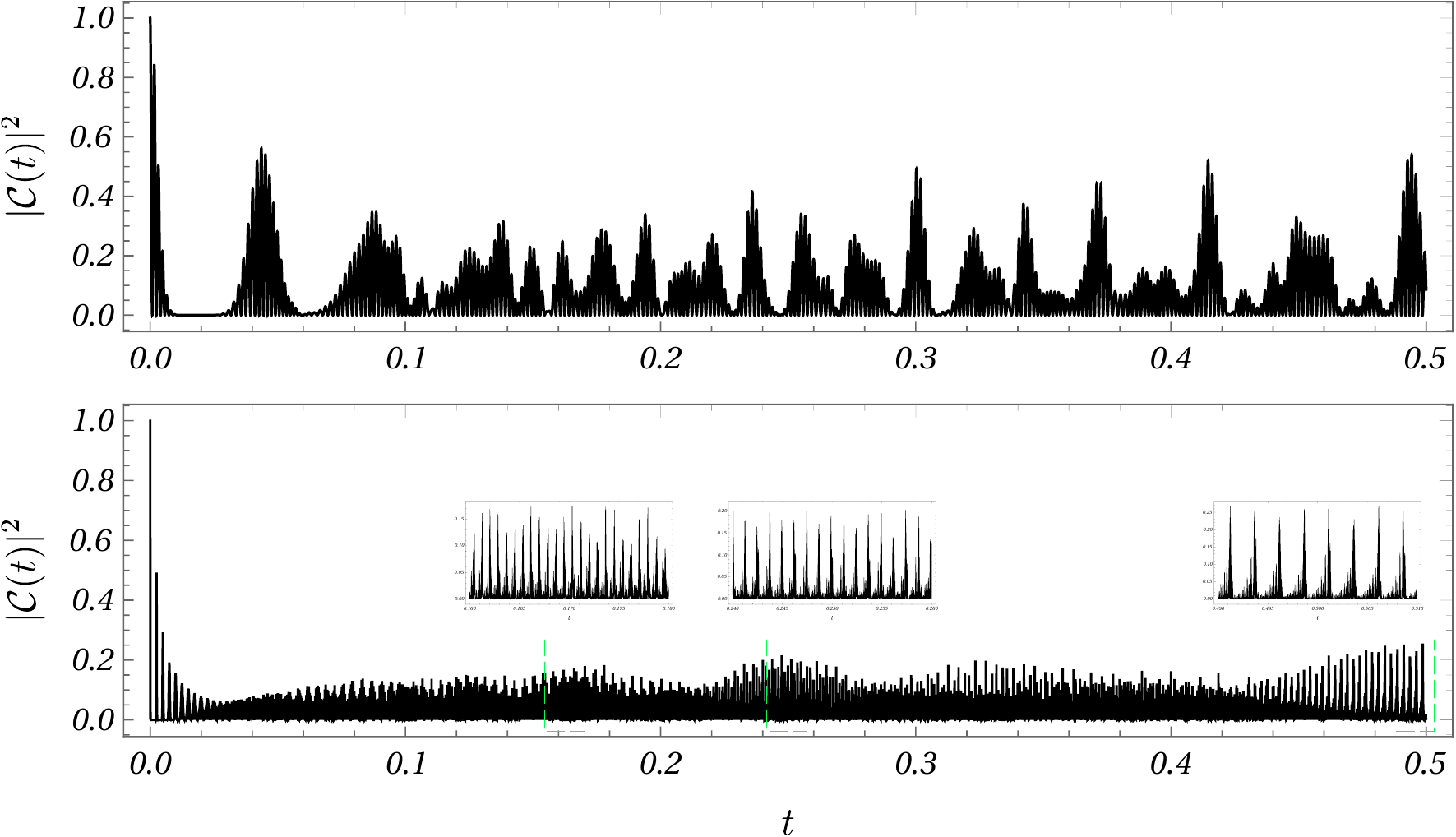} 
 \\
 \caption{{Revival pattern distortion of the survival probability in a strong magnetic field, which corresponds to setting one-particle parameters (see Eq.~\eqref{parameters}) to $A_{n_0}= 0, B_{n_0}= 1$, and $\eta_{n_0} =1/2 $ as in figure \eqref{autocorrelation1}. The horizontal axis is given in units of $T_2 ^{-1}$ where $T_2 = 3.7 \times 10^2,\,2.5 \times 10^4$ was calculated with Eq.~\eqref{timescales} for $a=5,\,20$, respectively, from top to bottom row. Insets on last plot confirm that local periodicities are preserved, with a fractional revival pattern occurring at $t/T_2 = 1/6,\, 1/4,\, 1/2$ with periods of $T_1/3,\, T_1 /2$, and $T_1$, respectively, where $T_1/T_2 \approx 4 \times 10^{-2} \, (a=5),\, 3 \times 10^{-3} \,(a=20)$. Fig.~\eqref{t1scale} depicts the evolution at the shorter time scale $T_1$, which is useful to compare the smallness of these ratios. }}
 \label{autocorrelation2}
\end{figure}

{\bf Fractional revivals also occur at these new time scales\footnote{One notices that a shorter revival scale has also been observed in the infinite well problem for the Schr\"{o}dinger equation \cite{Aronstein}, due to reflection symmetry. The half revival is equivalent to the full revival for even eigenstates only.}, which can be observed at the scale of $T_2$ in Fig.~\eqref{autocorrelation2}, where the initial oscillation with period $T_1$ was just described. They can be defined generally as the time scales at which correlated copies of the initial state oscillate with fractions of the initial period. The signature of fractional revivals can thus be identified in the survival probability as an oscillation which is faster than the usual revival. In fact, fractional revivals occur at rational multiples of the revival time, which also determine the local periodicity of the quantum state. Explicitly, the periodicity is given by fractions of the form } $T_1/q$ (odd $q$), at $t = (p/q)T_2$, where $p,q$ are mutually prime \cite{Robinett}. If $q$ is even, the local periodicity is $2T_1/q$. For instance, the most evident fractional revivals are half ($q=2$) and quarter ($q=4$) revivals. The former can be observed at $t= T_2/2$, when the state has a local periodicity $T_1$, and the latter can be observed at $t=T_2/4$ and $t=3T_2/4$, with periodicity $T_1/2$. One notices that peaks observed close to the quarter revival at $t= T_2/4$ and half revival at $t= T_2/2$ do not have a regular amplitude in Fig.~\eqref{autocorrelation2} for $a=5$, even if they oscillate with the predicted periods. The reason is twofold. First, the pattern is distorted because the interval between both fractional revivals is short, and thus they interfere. Second, the energy spread is significant for the average Landau level considered. Therefore, long time scales are less defined in strong magnetic fields.

As more eigenstates are excited, more fractional revivals can be resolved. One additional example was included in the last plot for $a=20$, where the fractional revival at $t=(1/6)T_2$ is observed with periodicity of $T_1/3$, as predicted. However, peaks in the survival probability become less pronounced with increasing spread of excited eigenfunctions (cf. Fig.~\ref{powerspectrum}). This also means that detecting fractional revivals exactly at multiples of the classical period, or multiples of its fractions \cite{Robinett, Krueckl}, is not possible here. 

{\bf Finally, it is also possible to detect the super revival time, $T_3$, when full revival states are not sufficiently close to the initial one. The qualitative pattern is similar to the revival time at longer scales. However, if higher order derivatives of the energy eigenvalues in Eq.~\eqref{enexpansion} vanish, then these time scales become infinite.  For a relativistic dispersion, all orders could be observed in principle, as long as the energy distribution of the initial state does not spread significantly (cf. Fig~\ref{powerspectrum}). The quantum state oscillates at fractions of $T_3$, as explicitly shown below. }The super revival time is depicted in Fig.~\eqref{super} for cat states with $a=5$, which were observed by moving to a frame with $k_z \neq 0$. The predicted value for the super revival period for wave packets is eight times greater (cf. Eq.~\eqref{timescales}) than $T_3$ for cat states, typically demanding a long time propagation scheme. Similar to the revival scale, the super revival modulates all the higher frequency oscillations described previously. There is a slight difference in that the local periodicity is not given by fractions of $T_2$ only, but depends on both $T_1$ and $T_2$ if the system is periodic. For a Dirac-like dispersion considered here, this subtle correction is not relevant, and the system initially oscillates with well-defined periods of $T_2/2 \approx 0.01 T_3$ (half revivals) for roughly six periods, recovering well-defined oscillations at $t/T_3 \approx 0.08, 0.16$, and $0.33$, with corresponding periodicity of $T_2/4,\, T_2/2$, and $T_2/2$. In fact, the first two periods have been predicted for Rydberg atoms \cite{Robert1,Robert2}, while the last one was expected to be $T_2$. 

For Dirac cat states studied here, super revivals can be observed if $A_{n_0}/B_{n_0} \gtrsim 2$ (cf. Eq.~\eqref{parameters}), with the survival probability almost returning to unity, which means that even if there are many eigenstates contributing to the quantum superposition, the eigenvalues spacing can be adjusted, and the same is valid by increasing the mass parameter. For greater values of this ratio, peaks in the survival probability are more noticeable, but time scales increase considerably. Conversely, if $A_{n_0}/B_{n_0} $ is not high enough, the revival structure repeats itself at the scale of $T_2$, which stems from other time scales that become relevant in the energy expansion. It is possible to see that the uneven spacing of relativistic Landau levels distort the revival structure, and the revival pattern becomes well-defined for weak magnetic fields. 
\begin{figure}
 \centering
 \includegraphics[width=1. \textwidth]{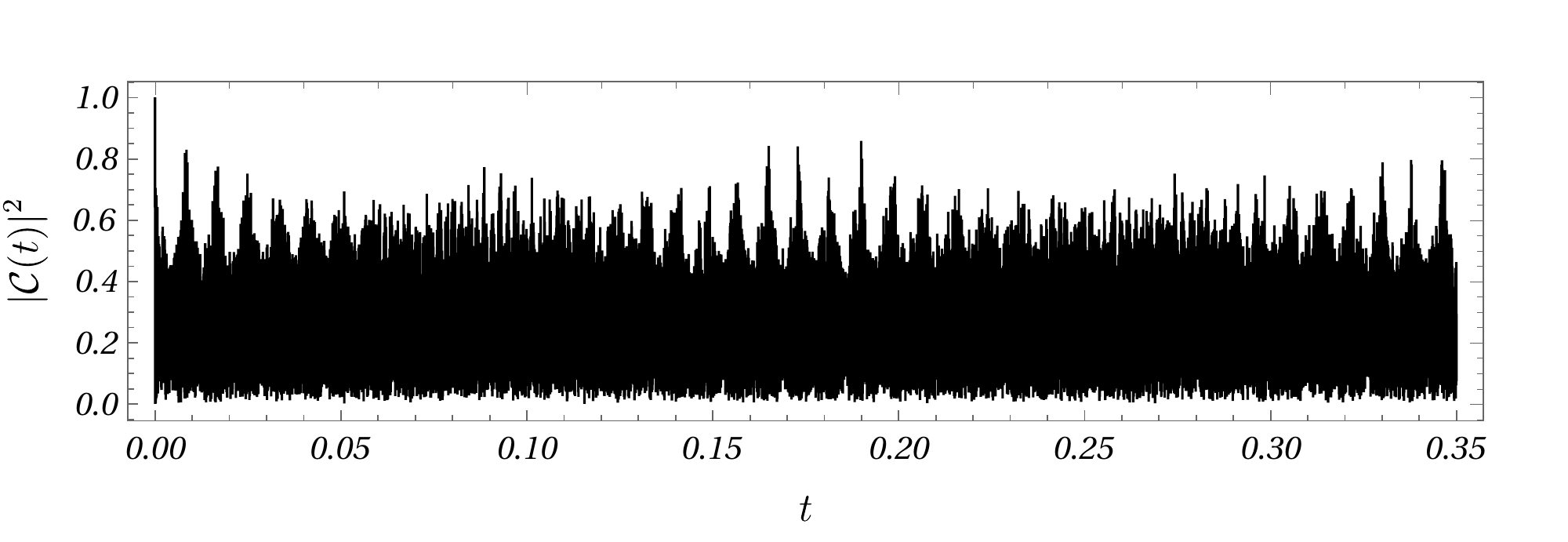}
 \caption{Survival probability for a Dirac cat state with $a=5$ at the $T_3 = 2.7 \times 10^5$ (super revival) time scale. One-particle parameters are set to $A_{n_0}/B_{n_0} = 2.04$ and $\eta_{n_0} =1/2$ (cf. Eq.~\eqref{parameters}). The time axis is given in units of $T_3 ^{-1}$, with smaller scales ratios given by  Eq.~\eqref{timescales}, i.e., $T_2 / T_3 \approx \num{1.6e-2} $.}
 \label{super}
\end{figure}
To sum up, this section has described the long time behavior of the survival probability, exhibiting collapse and revivals in a very correlated pattern between the excited eigenstates. The intricate state evolution and the oscillation frequencies are fully encoded in the energy distribution as previously described by the spectral function. However, since there are many excited eigenfunctions for cat states, the computation of the in-phase and out-of-phase copies of the initial state at fractional revivals is not practical. 
In the next section, an alternative description of the revival structure will be also given by computing useful quantum operators. Since the typical motion in terms of the position operator is not appropriate here, the localized probability density picture turns out to be useful, where it is possible to detect the oscillation of the two Gaussian states in the $(s,t)$ plane. Moreover, spinor matrix operators that exhibit revivals can be used to describe the temporal evolution of the intrinsic quantum information profile of cat states.

\section{Quantum Correlation Revivals}\label{sec4}
Following previous results of the quantum informational approach to spinor fields in frame-dependent calculations \cite{BernardiniEPJP,PRA2021}, one can inquire into other quantum operators that exhibit the revival structure. They are obtained from the spinor decomposition of the matrix-valued Wigner function in terms of the 16 independent generators of the Clifford Algebra. Using the Dirac representation, for which the gamma matrices are given by $\gamma_{0} = \beta$, $\gamma_{j} = \beta\alpha_j$, $\{\gamma_{\mu},\gamma_{5}\} = 0$, and $\sigma_{\mu\nu} = (i/2)[\gamma_{\mu},\gamma_{\nu}]$, the decomposition of the covariant Wigner function yields \cite{Weickgenannt}
\begin{equation}\label{51}
W(\{q\}) \equiv\mathcal{S}(\{q\})+
i\,\gamma_{5}\,{\Pi}(\{q\})+
\gamma_{\mu}\,\mathcal{V}^{\mu}(\{q\})+
\gamma_{\mu}\gamma_{5}\,\mathcal{C}^{\mu}(\{q\})+
\frac{1}{2}\sigma_{\mu\nu}\mathcal{T}^{\mu\nu}(\{q\}),
\end{equation}
with $\{q\}\equiv \{\mathbf{x},\,\mathbf{k};\,t\}$. Multiplying the left-hand side by the corresponding generator that appears in front of each term and, tracing over spinor indices, scalar, pseudo-scalar, vector, axial-vector, and anti-symmetric tensor contributions under Lorentz transformation are all correspondently identified, that is, 
\begin{eqnarray}
\mathcal{S}(\{q\}) &=&\frac{1}{4}Tr[W(\{q\})],\label{55a}\\
{\Pi}(\{q\})&=&-\frac{i}{4}Tr[\gamma_{5}\,W(\{q\})], \label{55b}\\
\mathcal{V}^{\mu}(\{q\})&=&\frac{1}{4}Tr[\gamma^{\mu}\,W(\{q\})],\label{55c}\\
\mathcal{C}^{\mu}(\{q\})&=&\frac{1}{4}Tr[\gamma_{5}\gamma^{\mu}\,W(\{q\})],\label{55d}\\
\mathcal{T}^{\mu\nu}(\{q\})=-\mathcal{T}^{\nu\mu}(\{q\})&=&\frac{1}{4}Tr[\sigma^{\mu\nu}\,W(\{q\})], 
\label{55e}\end{eqnarray}
where $Tr[...]$ refers to the trace over spinor indices. 

To show how a more convenient Wigner function can be obtained directly for spinor configurations in a fixed frame, one introduces briefly the 4-vector notation for $x^\mu=(t, \mathbf {x}), u^\mu = (\tau, \mathbf {x}), k^\mu = (k_0, \mathbf {k})$. A general spinor $\phi_\lambda (x)$ can be expanded in the eigenfunction basis as\footnote{
For now, arbitrary quantum superpositions can be considered, but the result will be applied to the problem considered here. {\bf Also, the momentum dependence has been omitted for notation compactness. }}
\begin{equation}
\phi_\lambda (x + u ) = \sum _{j} \psi_{\lambda,j} (\mathbf{x} + \mathbf{u})\exp[-i k_{0,j} (t + \tau)]
\end{equation}
where $j$ stand for the $j$-th spinor in the superposition for a particular orthonormal basis. The so-called equal-time Wigner function can then be computed as \cite{PRA2021}
\begin{eqnarray}
\omega_{ \xi \lambda} (\mathbf{x},\mathbf{k};t)&=& \int ^{+\infty} _{-\infty} \hspace{-1em} d \mathcal{E} \, W_{\lambda \xi} (x,k) \nonumber \\
&=& \pi^{-1} \sum _{j,m} \exp[ i(k_{0,j} - k_{0,m} ) t] \int d\tau \int ^{+\infty} _{-\infty} \hspace{-1em} d \mathcal{E} \exp[ -i( 2\mathcal{E} - k_{0,j} - k_{0,m} ) \tau] \nonumber \\
&& \quad \times \quad \pi ^{-3} \int d^3 \mathbf{u} \exp[2i \mathbf{k}. \mathbf{u}] \bar{\psi}_{\lambda,j}(\mathbf{x} - \mathbf{u})\psi_{\xi,m}(\mathbf{x} + \mathbf{u}) \nonumber \\
&=& \pi^{-3} \sum _{j,m} \exp[ i(k_{0,j} - k_{0,m}) t] \int d^3 \mathbf{u} \exp[2i \mathbf{k}. \mathbf{u}] \bar{\psi}_{\lambda,j}(\mathbf{x} - \mathbf{u})\psi_{\xi,m}(\mathbf{x} + \mathbf{u}).\label{stationarywignerfunction}
\end{eqnarray}
It supports a decomposition over Hermitian generators \cite{Zhuang},
 \begin{eqnarray}
 \label{V4}
 (\omega \gamma_0 ) (\{q\}) =
 && \!\!\!\!\!\!
 1/4 \Bigl[ f_0(\{q\})+ \gamma_5 f_1(\{q\})
 -i \gamma_0 \gamma_5 f_2(\{q\})
 + \gamma_0 f_3(\{q\})
 \\
 && - \gamma_0 \gamma_5 \mathbf{\gamma \cdot}{\bf g}_0(\{q\})
 + \gamma_0 \mathbf{\gamma \cdot}{\bf g}_1(\{q\})
 - i\mathbf{\gamma \cdot}{\bf g}_2(\{q\})
 - \gamma_5 \mathbf{\gamma \cdot}{\bf g}_3(\{q\})\Bigr]\, ,
 \nonumber
 \end{eqnarray}
where the phase-space densities in Eqs.~(\ref{55a})-(\ref{55e}) were split into time ($f_0,f_1, f_2,f_3$) and spatial (${\bf g}_0,{\bf g}_1,{\bf g}_2,{\bf g}_3$) components, with ${\bf g}_i = (g_i ^x, g_i ^y, g_i ^z)$, and spinorial indices are implied in the left-hand side. One notices that $\omega \gamma_0$ is also Hermitian, and all phase-space densities are real. 

The Hermitian decomposition is a more natural choice for frame-dependent calculations in a magnetic field, since phase-space expectation values of the physical densities can be obtained directly from the matrix elements of the corresponding Hermitian operator for pure states. In fact, by decomposing the observables in a fixed frame, one can evaluate their temporal evolution from initial conditions.

For instance, in the quantum information setting, the (charge) normalization can be imposed as 
\begin{eqnarray} \label{normalization}
\int d^3 \mathbf{x} \, \bigg ( \rho(\mathbf{x}) \bigg ) &=& \int d^3 \mathbf{x} \, \bigg ( \int d^3 \mathbf{k} \, Tr[ \omega \gamma_0 ] \bigg ) \nonumber \\
&=&1/4 \int d^3 \mathbf{x} \, \int d^3 \mathbf{k} \, f_0(\mathbf{x},\,\mathbf{k};\,t ) = 1,
\end{eqnarray}
for any density matrix that describes normalized probability distributions. Of the remaining phase-space densities, a few of them have an evident meaning. The spatial components of the Dirac current can be written as
\begin{eqnarray}\label{diraccurrent}
\int d^3 \mathbf{x} \, \mathbf{j}(\mathbf{x}) &=& \int d^3 \mathbf{x} \, \bar{\psi} \mathbf{\gamma} \psi = \int d^3 \mathbf{x} \, \bigg ( \int d^3 \mathbf{k} \, Tr[ \omega \mathbf{\gamma}] \bigg ) \nonumber \\
&=& 1/4 \int d^3 \mathbf{x} \, \int d^3 \mathbf{k} \, {\bf g}_1(\mathbf{x},\,\mathbf{k};\,t ),
\end{eqnarray}
and the spatial components of the pseudo-vector contribution can be written as a spin density, with 
\begin{eqnarray}\label{spindensity}
&&\int d^3 \mathbf{x} \, \mathbf{j} _{\gamma_5}(\mathbf{x}) = - \int d^3 \mathbf{x} \, \bar{\psi} \gamma_ 5 \bm{\gamma} \psi = \int d^3 \mathbf{x} \, \psi^\dagger \mathbf{\Sigma} \psi \nonumber \\
&=& - \int d^3 \mathbf{x} \, \bigg ( \int d^3 \mathbf{k} \, Tr[ \omega \gamma_5 \mathbf{\gamma}] \bigg ) = 1/4 \int d^3 \mathbf{x} \, \int d^3 \mathbf{k} \, {\bf g}_0(\mathbf{x},\,\mathbf{k};\,t ),
\end{eqnarray}
where $\gamma_0 \gamma_5 \bm{\gamma} = \mathbf{\Sigma} = \mbox{diag}\{\bm{\sigma,\sigma}\}$. The vector current describes the temporal evolution of the velocity operator, whereas the pseudo-vector current describes the temporal evolution of the spin operator. 

Once the charge normalization in Eq.~\eqref{normalization} is imposed in the basis of spinors, no general rule can be obtained for other components in the spinor decomposition. {\bf For a pure state, any phase-space averaged density can be calculated from the coordinate representation as the expectation value of the corresponding Hermitian generator in Eq.~\eqref{V4},} since they span the vector space of $4 \times 4$ matrices. This means that observables can be computed without evaluating the full matrix-valued Wigner function. One now specializes the result to ensembles of cat states.

\subsection{Probability density picture
}

To compute the temporal evolution of operators from Eq.~\eqref{V4} for cat states, one needs to evaluate the matrix elements between excited eigenstates in the form 
\begin{equation} \label{selectionrule}
 \langle u_{n,r} ^{\nu} \vert \Gamma \vert u_{m,r'} ^{\nu'} \rangle,
\end{equation}
where $\Gamma$ denotes the Hermitian generators in the spinor decomposition. The symmetries of cat states simplify the computation, since it is possible to show that matrix elements for constant matrices between Landau levels with $m\neq n$ are always zero. The only non-vanishing terms always correspond to $n=m$, i.e., the expectation values are computed between Landau levels with the same quantum number $n$. In contrast, a typical kicked Gaussian state \cite{Krueckl,Rusin}, localized in space with an initial momentum, would select $n= m\pm1$, if $\Gamma$ is written as a non-diagonal block matrix. This is also observed for the expectation values of in-plane position operators. Therefore, non-vanishing expectation values never mix the expansion coefficients in the superposition defined by Eqs.~(\ref{levela0})-(\ref{levela2}) for distinct HO quantum numbers. 

Consequently, the computation of the matrix elements can be straightforwardly obtained. For compactness of notation, the momentum averaged densities are denoted with the same notation as the scalar functions in phase space, with arguments replaced as $ \{\mathbf{x},\,\mathbf{k};\,t\} \rightarrow(s,t)$, where it is assumed that expectation values are taken between states with the same quantum numbers (continuous variable space), and thus the additional spatial degrees of freedom are constant plane waves that do not affect the result. The functions obtained are also described by scalar functions with spatial parity symmetry, as it will be shown shortly. 

To exemplify the selection rule for symmetric cat states, the conservation of the time component of the Dirac current (see Eq.~\eqref{normalization}) is obtained explicitly from the normalized probability density
\begin{eqnarray}
\frac{f_0 (s,t)}{4} &=& \frac{1}{\cosh{(a^2/2)}} \sum_{n,m} \frac{(a/\sqrt{2})^{n+m}}{\sqrt{n! m!}}\Bigg[ \mathcal{F} _{n-1} (s) \mathcal{F} _{m-1} (s) \cos((E_n - E_m)t) - \nonumber\\
&& \sum \sin(E_n t) \sin(E_m t) B_n B_m \eta_n \eta_m \bigg( \mathcal{F} _{n-1} (s) \mathcal{F} _{m-1} (s) - \mathcal{F} _{n} (s) \mathcal{F} _{m} (s) \bigg) \Bigg], \label{localprob}
\end{eqnarray}
with
\begin{eqnarray}
(1/4) \int \, ds f_0 (s,t) &=& \frac{1}{\cosh{(a^2/2)}} \sum_{n,m} \frac{(a/\sqrt{2})^{n+m}}{\sqrt{n! m!}} \Bigg[ \delta_{n,m} \cos((E_n - E_m)t) - \nonumber\\
&&\sum_{n,m} \sin(E_n t) \sin(E_m t) B_n B_m \eta_n \eta_m \bigg( \delta_{n,m} - \delta_{n,m} \bigg) \Bigg], \nonumber \\
&=&\frac{1}{\cosh{(a^2/2)}} \sum_n \frac{(a/\sqrt{2})^{2n}}{n!} = 1,
\end{eqnarray}
and $n,\,m$ running over even numbers. After integration, only $n=m$ terms survive. Hence, the computation of other operators is left to Appendix \ref{AppC} and shall henceforth be restricted to Landau levels with $n=m$.

The probability density in Eq.~\eqref{localprob} has parity symmetry around the origin $s=0$, which can be used to clarify the fractional revival structure of the survival probability function. First, the energy expansion around $E_{n_0}$ can be plug into the $\cos((E_n - E_m)t)$ term, 
\begin{equation}
 (E_n - E_m)\approx E'(n_0)(n-n_0) + \frac{E''(n_0)}{2}(n-n_0)^2 + ...
\end{equation}
which describes the lower frequency oscillations. Since the expansion coefficients are always even, the doubling in each term is preserved. On the other hand, terms proportional to $\sin(E_n t)\sin(E_m t)$ indicate that the rapid oscillation with frequency $E_{n_0}$ is preserved unless $B_{n_0} \eta_{n_0} \approx 0$, in the weak field limit. Thus, all time scales can be detected in the probability density, and the contributions of Hermite polynomials of even and odd parities mean that symmetric and antisymmetric states exhibit the same oscillation pattern.

\begin{figure}
 \centering
 \includegraphics[width=0.85\textwidth]{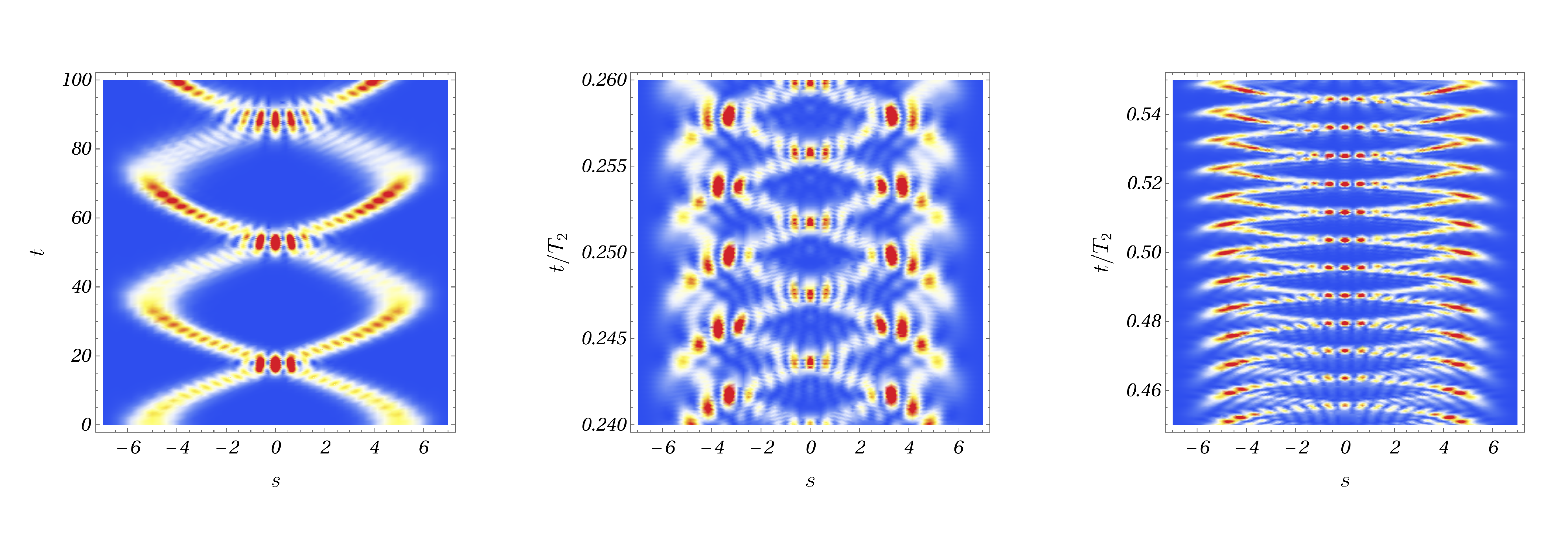}
 \caption{Contour plot of the probability density $f_0 (s,t)$ (\ref{localprob}) for a Dirac cat state with a color scheme that identifies the maximum probability density (red). The state initially reforms twice with a period $T_1 = 35$ (left) but oscillates afterwards with fractions of this period at fractional revivals. For instance, at $t/T_2 = 1/4$ (middle), with a period of $T_1/2$, corresponding to a quarter revival, and $T_1$ again at $t/T_2 = 1/2$, a half revival (right). Hamiltonian parameters are set as in Fig.~\eqref{super}, $A_{n_0}/B_{n_0} = 2.04$ and $\eta_{n_0} =1/2$ (cf. Eq.~\eqref{parameters}) for $a=5$, which yields $T_2 = 4 \times 10^3$ ($T_1 / T_2 \approx \num{8e-3} $).
 } 
 \label{contour}
\end{figure}

As one can see in Fig.~\eqref{contour}, the probability density describes the evolution of two Gaussian states initially localized at $s = \pm a$, each exhibiting the periodicity defined by the energy expansion in Eq.~\eqref{enexpansion}. The halving of the classical time then discussed for cat states in Eq.~\eqref{timescales} is observed because the two localized states swap positions and reform an identical state, symmetric or antisymmetric. The initial motion of cat states thus seem to mimic quarter revival states of a parent localized state, which splits into spacial shifted copies of itself. Each copy oscillates with $2T_1$, the standard classical period. However, since they are indistinguishable, as shown by the survival probability, the cat state reforms itself at $t=T_1$, half the classical period. For the same reason, at half the revival time $t=T_2/2$, the initial state is recovered, and full and half revivals are indistinguishable. 

Considering that observables are described by averaged operators, the probability density picture works as a means of elucidating the quantum state temporal evolution. One now computes the averaged quantities that take into account the contribution of both Gaussian-localized states. 

\subsection{Entanglement and Mutual Information}

{\bf The tensor product structure of the Dirac representation evinces that the expectation value of operators is closely tied with the quantum information structure of fermions in a constant magnetic field. In fact, spinor operators can be regarded as two-qubit operators associated to the internal degrees of freedom of the $SU(2)\otimes SU(2)$ group structure \cite{MeuPRA} implied by the Dirac equation. Thus, it is expected that the time-dependence of these observables can affect the usual quantifiers of quantum correlations. }

The ensemble-averaged quantum operators can be obtained with a few algebraic manipulations as it was exemplified for the probability density and further discussed in Appendix \ref{AppC}. The obtained averaged values read
\footnotesize
\begin{eqnarray}
(1/4)\int \, ds \, g_1 ^z (s,t) &=& \langle \alpha_z \rangle _{CS} = \frac{-2M}{\cosh{(a^2/2)}} \sum _n \frac{(a^2/2)^{2n}}{(2n)!}\frac{\eta_{2n+1} A_{2n+1}}{ E_{2n+1}} \sin({E_{2n+1}t})^2 \label{eqaa1},\\
(1/4) \int \, ds \, g_0 ^z (s,t) &=& \langle \gamma_5 \alpha_z \rangle _{CS} = 1 - \frac{8}{\cosh{(a^2/2)}} \sum _n\frac{(a^2/2)^{2n}}{(2n)!} \eta_{2n+1} ^2 B_{2n+1}^2 \sin({E_{2n+1}t})^2 \label{eqaa2} ,\\ 
(1/4) \int \, ds \, g_2 ^z (s,t) &=& - \langle i \gamma_z \rangle _{CS} = \frac{2}{\cosh{(a^2/2)}} \sum _n\frac{(a^2/2)^{2n}}{(2n)!} \eta_{2n+1} A_{2n+1} \sin(2 {E_{2n+1}t}), \label{eqaa3}, \\
(1/4) \int \, ds \, g_3 ^z (s,t) &=& - \langle \gamma_5 \gamma_z \rangle _{CS} = 1 - \frac{8}{\cosh{(a^2/2)}} \sum _n\frac{(a^2/2)^{2n}}{(2n)!} \eta_{2n+1} ^2 A_{2n+1}^2 \sin({E_{2n+1}t})^2, \label{eqaa4} \\
(1/4) \int \, ds \, f_3 (s,t) &=& \langle \gamma_0 \rangle _{CS} = 1 - \frac{8}{\cosh{(a^2/2)}} \sum _n\frac{(a^2/2)^{2n}}{(2n)!} \eta_{2n+1} ^2 (A_{2n+1}^2 + B_{2n+1}^2) \sin({E_{2n+1}t})^2, \label{eqaa5}
\end{eqnarray}
\normalsize
with $\langle i \gamma_z \rangle _{CS} = \langle i \gamma_0 \gamma_5 \rangle _{CS}$ and $\langle \gamma_5 \rangle _{CS} = - \langle \alpha_z \rangle _{CS}$. The second column was written as the ensemble average evaluated for the symmetric Dirac cat state (CS) introduced in Eqs.~(\ref{levela0})-(\ref{levela2}). 

The first and second lines, Eqs.~\eqref{eqaa1} and \eqref{eqaa2}, correspond to the non-vanishing components, in the direction of the magnetic field, of the velocity and spin density operators, respectively. The perpendicular components, for instance, $\alpha_x$ and $\alpha_y$, vanish. This is a signature of the symmetric property of cat states, since a propagating quantum state in either direction would not be symmetric. In this sense, they are confined.

The phase-space averaged densities can be directly obtained from the expectation values in coordinate representation, since the quantum system is pure, and thus the computation in both cases simplifies to matrix elements of the corresponding quantum operator in the eigenfunction basis. These expectation values have a particular relevance when considering Dirac bispinors as spin-parity entangled qubits. Phase-space averaged quantum information quantifiers recently proposed in \cite{BernardiniEPJP,PRA2021} can be built explicitly from specific combinations of the functions above for a pure Dirac cat state. For instance, by rewriting $\langle \gamma_5 \gamma_0 \gamma_z \rangle $ as the spin density $ \langle \Sigma_z \rangle $ it is possible to show that
\begin{equation}
\big(1 + \langle \gamma_0 (t)\rangle\big)\left(\frac{1 - \langle \Sigma_z (t) \rangle }{2}\right) = \langle \mathcal{
C}^2 (t) \rangle ^{SP} _{x,k_x},
\label{conc}
\end{equation}
where the $CS$ index was suppressed. The right-hand side of Eq.~(\ref{conc}) corresponds
to the squared spin-parity quantum concurrence averaged in phase space, and the left-hand side is written as the product between functions of the intrinsic parity and spin operators, which confirms that, for a pure state, quantum states with well-defined spin and intrinsic parity can be written as a spin-parity separable spinor. This can be regarded as the temporal evolution of the intrinsic entanglement of a cat state. It follows that the zeros of the spin-parity quantum concurrence correspond to revivals of these quantum operators, starting from a spin-parity product state.

The revival structure of the survival probability for cat states thus allows one to predict the behavior of quantum correlations at long times. In particular, this can be applied to simulated relativistic Landau levels \cite{Rusin}. One relevant difference between mesoscopic states reported in \cite{Bermudez} is that they exhibit entanglement between spinorial and orbital degrees of freedom. Here, intrinsic entanglement quantifies correlations, also mediated by the HO basis, between the discrete degrees of freedom implied by the $SU(2)\otimes SU(2)$ group structure supported by the Dirac equation \cite{extfields}. Hence, averages of quantum operators need to take into account the normalization of the Hermite polynomials, regardless of the particular regime considered. This is important because by otherwise disregarding orthogonality relations, it is no longer possible to keep track of the purity constraint of a two-qubit system implied by Dirac bispinors in a magnetic field, as formally introduced in \cite{BernardiniEPJP}. 

Using similar arguments, the mutual information between phase-space and spin-parity Hilbert space\footnote{This is twice the linear phase-space entropy, which measures the degree of localization of a quantum state.} can also be expressed in terms of expectation values, 
\begin{eqnarray}
\langle \mathcal{
M} (t) \rangle _{x,k_x} ^{SP} = 2 &-& \frac{1}{2} \bigg( \big(1 + \langle \gamma_0 (t)\rangle\big)^2 + \big(\langle \gamma_0 \Sigma_z (t) \rangle +1\big)^2 + \nonumber \\
&&
\big(\langle\Sigma_z (t) \rangle -1\big)^2 - \langle i \gamma_z (t) \rangle^2 - 4 \langle \alpha_z (t) \rangle^2 \bigg),
\label{mutual}
\end{eqnarray}
where $\langle \gamma_5 \gamma_z \rangle $ was written as $ \langle \gamma_0 \Sigma_z \rangle $ and one notices that all averaged quantum operators contribute to this quantifier\footnote{An alternative description of revivals in terms of position and momentum entropies has also been proposed in \cite{Romera4}.}. 

Considering that the connection between intrinsic quantum concurrence and phase-space information entropies has been investigated \cite{PRA2021}, the mutual information can be used to quantify the exceeding classical-like information with respect to the quantum concurrence. The expression is less intuitive than Eq.~\eqref{conc}, but one relevant point is that the explicit dependence on the kinetic terms, for instance, the vector current $\langle \alpha_z \rangle$, shows that even for weak magnetic fields, the mutual information quantifier does not vanish. Instead, the negative sign shows that it is responsible for classical-like correlations. 

Since the calculated mean values exhaust the spinor decomposition, it is no surprise that the elementary information of Dirac bispinors can be inferred from the mean value of quantum operators described above. This might be expected for a pure state, where the one-to-one correspondence between the spinor-valued wave function and its corresponding Wigner matrix is verified. However, this would still hold for mixed states. In this case, the revival structure could be detected in the temporal evolution of quantum operators obtained from the decomposition of the Wigner matrix, even without direct access to the spinor wave function. 

\begin{figure}
 \centering
 \includegraphics[width=\textwidth]{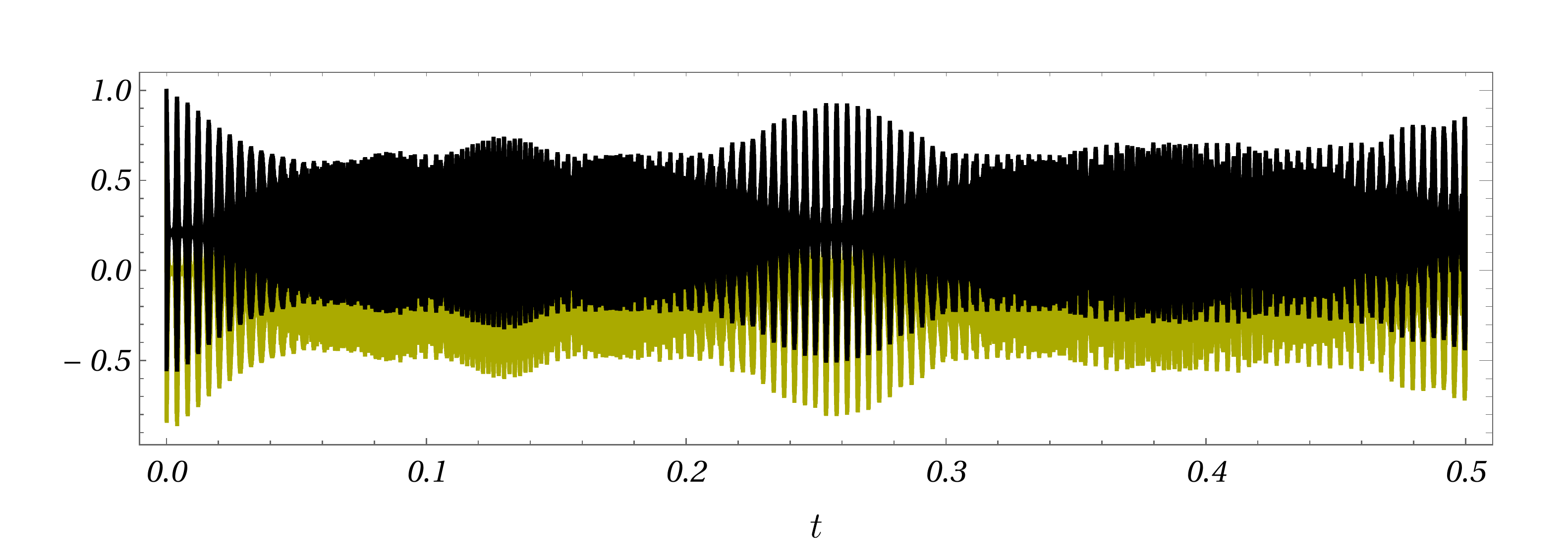}
 \caption{Revivals in the ensemble-averaged tensor components from Eqs.~(\ref{eqaa3})-(\ref{eqaa4}), $(1/4)\int_s g_2 ^z (s,t) $ (yellow, light gray) and $(1/4)\int_s g_3 ^z (s,t)$ with $B_{n_0}/A_{n_0} \approx 0.49 $, $\eta = 1/2$ (cf. Eq.~\eqref{parameters}), and $a=5$. The variable $t$ is scaled as $(T_2)^{-1}$, with $T_2 = 4.3 \times 10^3 $. Full and half revivals are observed at $t=0.5$ and $t=0.25$, respectively. Quarter revivals are observed at $t=0.125$ and $t=0.375$.}
 \label{observables0}
\end{figure}

To sum up, the selection rule for quantum operators reflect the symmetries of cat states. For operators that are not block diagonal matrices, their expectation values vanish. In contrast, operators that are written as block diagonal matrices have a non-trivial temporal evolution that also exhibits a fractional revival structure. The time scales detected in the survival probability are halved once again, which can be used to describe the long-time behavior of quantum correlations. Thus, whenever Dirac cat states reform, spin-parity degrees of freedom disentangle. As an illustration, two operators are depicted in Fig.~\eqref{observables0}, since the remaining expectation values in Eqs.~(\ref{eqaa1})-(\ref{eqaa5}) have a similar time-dependence pattern. The $T_2$ revival scale is shown to reproduce all features observed in the survival probability function, with the main difference that frequencies are doubled due to the $\sin ^2 (E_n t)$ terms. For instance, full revivals are observed at $t/T_2 = 1/2$, instead of $t=T_2$. When the energy is dominated by the kinetic term $k_z$ (cf. Eq.~\eqref{parameters}), the revival structure is nearly exact, since energy levels are almost evenly spaced. In this regime, quantum concurrence vanishes and the mutual information is only associated to classical correlations. 

The low-field regime corresponds to more accurate revival patterns, when mostly classical-like correlations are relevant. As a matter of fact, any revival partially reforms the initial state, which is spin-parity separable. Therefore, Dirac Hamiltonian parameters can be tuned to adjust to particular regimes for relativistic Landau levels considered, allowing a control over the revival structure of cat states and the corresponding correlation profile over long periods. 

\section{Conclusions}

Reporting about the intrinsic correlation structure of Dirac bispinors driven by the $SU(2) \otimes SU(2)$ spin-parity Hilbert spaces \cite{BernardiniEPJP,PRA2021}, and constrained by confining potentials, the long-time evolution of Dirac cat states with a localized energy distribution spread out in coordinate space was examined. Following the parity invariance of the Dirac equation, the summarized result here noticed is that cat states have a unique temporal evolution signature, which depends simultaneously on the initial conditions of the spinor degrees of freedom and on the symmetry associated to continuous variables, introduced by the electromagnetic confining potential. 

Gaussian localized Dirac cat states were obtained from spin and parity considerations, and their energy spectral function was investigated.
On one hand, energy- and position-localized quantum states were shown to exhibit a semiclassical behavior for high enough energy eigenvalues, at least for short times. On the other hand, symmetric states were shown to be not necessarily localized in coordinate space. Nevertheless, the established framework for wave packet dynamics allowed the evaluation of the intricate time behavior of these quantum superpositions. It was shown that cat states mimic a fractional revival structure in the sense they initially oscillate with half the classical periodicity, which can be predicted by the energy expansion around a mean eigenvalue for equivalent HO quantum numbers with the same parity. In fact, oscillation periods were re-scaled by powers of 2 when compared to wave packets. Since the expansion coefficients were expressed analytically, the phases acquired by each eigenstate were identified and could be controlled for long times. The conditions for which the revival and super revival scales are observed were also obtained in terms of the one-particle parameters. These were computed in natural units, allowing one to specialize them to weak- and high-field limit for a massive particle in any particular frame, or even to the ultra-relativistic limit, which is, for instance, supported by the mapping of relativistic Landau levels to quantum optic Hamiltonians.

As an extension, the connection between the survival probability and physical observables related to $SU(2) \otimes SU(2)$ spin-parity information quantifiers was investigated. The associated operators described the temporal evolution of the quantum information encoded in Dirac bispinors. For a Gaussian cat state ensemble, selection rules were significantly simplified, since principal quantum numbers always have the same parity. For the same reason, the momentum-averaged densities obtained from the Wigner function are also parity-defined, for which the spin density and the intrinsic parity operators are quantifiers related to the quantum entanglement of pure Dirac cat states. These operators have worked as a probe for the fractional revival structure depicted here.

To conclude, in the same sense that the energy localization of a quantum state has a straightforward correspondence with its temporal evolution, the addressed symmetric states have also exhibited a well-defined temporal behavior at long time scales, with an unavoidable revival structure identified from its associated unitary evolution. As mentioned, this has worked as a probe for controllable dynamical features in systems with a discrete energy spectrum. In particular, since the survival probability and other observables have been computed exactly in terms of the Dirac Hamiltonian parameters, these may be considered in mapping low-energy Dirac systems which engender quantum states through confining potentials with physically accessible scales. 
In addition, since the parity invariance is a general feature of the Dirac equation, and ensembles of cat states could also be prepared in other Hamiltonians with confining potentials of physical interest \cite{Sheng2}, the framework here discussed can be considered in several theoretical platforms, which include, for instance, the Dirac oscillator \cite{Mandal}, or external potentials that exhibit some correspondence with Schr{\"o}dinger-like systems \cite{Garcia}, even if a general rule for intrinsic correlations of Dirac cat states in position-dependent interactions is still incipient.
Finally, the initial conditions for cat states have been shown to be too strict, and most interactions that can be treated in a perturbation scheme would introduce additional selection rules that mix symmetric and antisymmetric states, thus destroying these fragile quantum superposition patterns. This opens up possible windows to interpret mixing as a driver of quantum and classical information dissipation. 

\vspace{.5 cm}
{\em Acknowledgments -- The work of A.E.B. is supported by the Brazilian Agencies FAPESP (Grant No. 2023/00392-8) and CNPq (Grant No. 301485/2022-4). The work of C.F.S. is supported by the Brazilian Agency Capes (Grant No. 88887.499837/2020-00).}

\appendix
\section{Intrinsic and extrinsic entanglement}
\label{AppA}
The spin-parity entanglement exhibited by Dirac spinors, solutions of the Dirac equation, is interpreted as intrinsic, or intraparticle, entanglement \cite{MeuPRA}. As it shall be explained in the following, this is different from entanglement between degrees of freedom associated to different particle states, as it should be the polarization (extrinsic) entanglement between different electrons.
The intrinsic entanglement is encoded in internal degrees of freedom of a single particle. An example resides in the framework of neutron interferometry: the spin state of the particle and a quantum state associated to different possible paths between the source and the measure apparatus can be entangled \cite{Neut01,Neut02}. Given that neutron states can be manipulated and measured, spin-path entanglement is a measurable output \cite{Neut01} used, for instance, in the discussion of Bell's inequality \cite{Neut02}. More relevantly, such a spin-path intrinsic (or intraparticle) entanglement can be suitably transferred to extrinsic (or interparticle) entanglement \cite{Neut03,PRA2018}. Other configuration can be set up from quantum optics: a single particle intrinsic entanglement can be encoded by single photons through different degrees of freedom, such as polarization and either orbital angular momentum \cite{photon01} or transverse spatial degree of freedom \cite{photon02}, and even other interferometry variables \cite{photon03}. In that case, several quantum information protocols were engineered to as to include the features of such an intraparticle entanglement in photon systems \cite{photon04, photon05}.

More specifically concerning Dirac-like equation structures, {\em lattice-layer} intrinsic entanglement structures of Bernal stacked bilayer graphene were also obtained for the quantum system described by a tight-binding Hamiltonian through a suitable correspondence with the {\em parity-spin} $SU(2)\otimes SU(2)$ structure of a Dirac Hamiltonian \cite{MeuPRB}.
In this case, the {\em lattice-layer} two qubit basis can be mapped into the Dirac {\em parity-spin} degrees of freedom as to support the interpretation of the bilayer graphene eigenstates as intrinsically entangled ones in a {\em lattice-layer} basis.

Finally, in the same scope, two-qubit states of a four-level trapped ion quantum system with ionic states driven by Jaynes-Cummings Hamiltonians can also have interactions mapped onto a $SU(2)\otimes SU(2)$ group structure. Again, such internal degrees of freedom corresponding to intrinsic parity and spin polarization work as a mapping platform for computing the quantum entanglement between the internal quantum subsystems which define two-qubit ionic states \cite{MeuPRA}.

In particular, to provide a more technical explanation for the intrinsic entanglement in the context of the $SU(2)\otimes SU(2)$ group structure, one should turn back to the grounds of the definitions of the representations of $SU(2)\otimes SU(2)$ as a subset of the $SL(2,\mathbb{C})\otimes SL(2,\mathbb{C})$ group.

In terms of the Lie algebras and Lie groups, the representations of $sl(2,\mathbb{C})\oplus sl(2,\mathbb{C})$ algebra, the Lie algebra of the $SL(2,\mathbb{C})\otimes SL(2,\mathbb{C})$ group, are irreducible: they correspond to tensor products between linear complex representations of $sl(2,\mathbb{C})$. Hence, unitary irreducible representations of $SU(2)\otimes SU(2)$ are tensor products between unitary representations of $SU(2)$, and one has a one-to-one correspondence with the group $SL(2,\mathbb{C})\otimes SL(2,\mathbb{C})$ and the same for the corresponding algebra.

In this case, it should be pointed out that {\em inequivalent representations} of $SU(2) \otimes SU(2)$ follows from the mentioned one-to-one correspondences. Inequivalent representations of $SU(2) \otimes SU(2)$ do not correspond to all the representations of $SL(2,\mathbb{C})\otimes SL(2,\mathbb{C})$ (and, consequently, to all the proper Lorentz transformations that compose the $SO(3,1)$ group). They just corresponds to a subset of $SO(4) \equiv SO(3)\otimes SO(3)$.
One thus may choose at least two {\em inequivalent} subsets of $SU(2)$ generators, such that $SU(2)\otimes SU(2) \subset SL(2,\mathbb{C})\otimes SL(2,\mathbb{C})$, with each generator having its own irreducible representations ($irrep$) symbolically described by $irrep (su_{\xi}(2)\oplus su_{\chi}(2))$.

Turning back to our arena, a {\em spinor} $\xi$ described by $(\frac{1}{2},\,0)$ transforms as a {\em doublet} (fundamental representation of $SU_{\xi}(2)$), and as a singlet (``transparent'' to transformations of the $SU_{\chi} (2)$). From the notation $(\mbf{dim}(SU_{\xi}(2)),\mbf{dim}(SU_{\chi}(2)))$, the {\em spinor} $\xi$ is an object identified by $(\mbf{2},\mbf{1})$, as well as the {\em spinor} $\chi$ is identified by $(0,\,\frac{1}{2})$. Both transform as a {\em singlet} of $SU_{\xi}(2)$ and as a {\em doublet} of $SU_{\chi}(2)$, respectively.

Additional elementary representations of $SL(2,\mathbb{C})$ can be identified by: $(\mbf{1},\mbf{1})$ -- a {\em scalar} or {\em singlet}, with angular momentum projection $j = 0$; $(\mbf{2},\mbf{1})$ -- a {\em left-handed spinor} $(\frac{1}{2},\,0)$, with angular momentum projection $j = 1/2$; $(\mbf{1},\mbf{2})$ -- a {\em right-handed spinor} $(0,\,\frac{1}{2})$, with angular momentum projection $j = 1/2$; and $(\mbf{2},\mbf{2})$ -- a {\em vector/doublet}, with angular momentum projection $j = 0$ and $j = 1$. Such elementary representations can be manipulated in order to give
$ (\mbf{1},\mbf{2}) \otimes (\mbf{1},\mbf{2}) \equiv (\mbf{1},\mbf{1}) \oplus (\mbf{1},\mbf{3}),$,
a representation that composes Lorentz tensors like
\begin{equation}
C_{\alpha\beta}\bb{\mt{x}} = \epsilon_{\alpha\beta} D\bb{\mt{x}} + G_{\alpha\beta}\bb{\mt{x}},
\end{equation}
where $D\bb{\mt{x}}$ is a scalar, and $G_{\alpha\beta} = G_{\beta\alpha}$ is totally symmetric,
$ (\mbf{2},\mbf{1}) \otimes (\mbf{1},\mbf{2}) \equiv (\mbf{2},\mbf{2})$,
such that
$ (\mbf{2},\mbf{2}) \otimes (\mbf{2},\mbf{2}) \equiv (\mbf{1},\mbf{1}) \oplus (\mbf{1},\mbf{3}) \oplus (\mbf{3},\mbf{1}) \oplus (\mbf{3},\mbf{3})
$,
which correspond to a decomposition into smaller {\em irreps} related to the Poincaré algebra classification \cite{extfields}.

The main point from the above construction is that the Dirac Hamiltonian dynamics can be discussed in terms of a group representation described by a direct product between two algebras which compose a subset of the group $SL(2,\mathbb{C})\otimes SL(2,\mathbb{C})$, the group $SU(2)\otimes SU(2)$. Out of the context of our work, Majorana, Weyl and some additional classes of spinor equations can also be driven by other subsets of $SL(2,\mathbb{C})\otimes SL(2,\mathbb{C})$.

In a much simpler context \cite{MeuPRA,extfields}, for the free particle Dirac Hamiltonian in the form of,
\begin{equation}
\hat{H}_{D}=\hat{\mbf{\alpha}}\cdot \hat{\mbf{p}}+m \hat{\beta},
\label{eqshamdirac}
\end{equation}
the Dirac state vectors are written as
$ \psi^{\dagger} \bb{\mt{x}} =\left(
\psi^{\dagger}_{\L}\bb{\mt{x}},
\psi^{\dagger}_{\R}\bb{\mt{x}}
\right)\equiv (\mbf{2},\mbf{2})$,
with {\em right-} and {\em left-handed} {\em spinors},
\begin{equation}
(\mbf{2},\mbf{1}) \equiv \psi _{\L}\bb{\mt{x}} =\left(
\begin{array}{c}
\psi _{\L\1}\bb{\mt{x}} \\
\psi _{\L\2}\bb{\mt{x}}
\end{array}
\right) ,\qquad (\mbf{1},\mbf{2}) \equiv\psi _{\R}\left( t\right) =\left(
\begin{array}{c}
\psi _{\R\1}\bb{\mt{x}} \\
\psi _{\R\2}\bb{\mt{x}}
\end{array}
\right).
\end{equation}

They are Dirac described as two {\em qubits} states encoded in a massive particle whose dynamics is represented by continuous variables (either $\mt{x}$ or \mt{p}). In the $\mbox{SU}(2)\otimes \mbox{SU}(2)$ framework, one has the free Hamiltonian given in terms of two-{\em qubit} operators, ${H}_{D}={{\sigma}}_{x}^{\left( 1\right) }\otimes \left(
{\bm p}\cdot {{{\mbox{\boldmath$\sigma$}}}}^{\left( 2\right) }\right) +m \,
{{\sigma} }_{z}^{\left( 1\right)}\otimes {I}^{(2)}_{2}$. The corresponding eigenstates are written as a sum of direct products describing \textit{spin-parity} intrinsically entangled states,
\begin{eqnarray}
\label{eqsB02}
\lefteqn{\left\vert \Psi ^{s}({\bm p},\,t)\right\rangle=
e^{i(-1)^{s}\,E_{p}\,t}\left\vert \psi ^{s}({\bm p})\right\rangle
}\\
&&= e^{i(-1)^{s}\,E_{p}\,t} N_{s}\left( p\right) \notag
 \left[ \left\vert
+\right\rangle _{1}\otimes \left\vert u({\bm p})\right\rangle _{2}+\left(
\frac{p}{E_{p}+(-1)^{s+1}m}\right) |-\rangle _{1}\,\otimes \left( {\bm p}
\cdot {{\mbox{\boldmath$\sigma$} }}^{\left( 2\right) }\left\vert u(\bm{p}
)\right\rangle _{2}\right) \right],
\end{eqnarray}
where $s = 0,\, 1$ stands for particle/antiparticle associated frequencies, and the spinor character is given by $\left\vert u(\mbf{p})\right\rangle _s$.
{\bf The intrinsic entanglement can be explained in the following terms \cite{MeuPRA,extfields}: $\left\vert u(\mbf{p})\right\rangle _{\2}$ is a {\em bi-spinor} the state that describes a fermion with a continuous momentum degree of freedom coupled to its {\em spin}, which describes a magnetic dipole moment in the case of a coupling with an external magnetic field.
The state (\ref{eqsB02}) is a superposition between parity eigenstates and therefore it does not exhibit a defined intrinsic parity.}
For the {\em qubit} $1$, the {\em doublets} $\left\vert + \right\rangle _{\1}$ e $ \vert - \rangle _{\1}$ are identified as the intrinsic parity eigenstates of the fermion.

The discriminating role of the parity can be better understand in terms of the total operator $\hat{P}$, $$\hat{P}\left( \left\vert \pm \right\rangle _{\1}\otimes \left\vert u(\mbf{p})\right\rangle _{\2}\right) =\pm \left( \left\vert \pm \right\rangle_{\1}\otimes \left\vert u(-\mbf{p})\right\rangle _{\2}\right),$$ which acts on the direct product $\left\vert \pm \right\rangle _{\1}\otimes \left\vert u(\mbf{p})\right\rangle_{\2}$. One indeed has the product of two operators: intrinsic parity, $\hat{P}^{int}$ (with two eigenvalues, $\hat{P}^{int}\left\vert \pm \right\rangle =\pm \left\vert\pm \right\rangle $) and spatial parity $\hat{P}^{s}$ (with $\hat{P}^{s}\varphi \left( \mbf{p}\right) =\varphi \left( -\mbf{p}\right) $), which acts on the continuous degrees of freedom.
{\bf All the analysis is constrained by the covariance properties related to the Poicar\'e algebra transformations.}
By applying $\hat{P}^{int}=\hat{\beta} =\hat{\sigma} _{z}^{\left( 1\right)}\otimes \hat{\mathbb{I}}_{\2}^{\left( 2\right) }$ to $\left\vert \psi ^{s}(\mbf{p},\,t)\right\rangle $, following Eq.~(\ref{eqsB02}), it follows that $ \hat{P} ^{\mi\1}=\hat{P}$, and the spatial parity resembles $\hat{P}^{int}$, as well as $\left( \hat{P}^{int}\right)^{\2}=\hat{\mathbb{I}}_{\2}^{\left( 1\right) }\otimes \hat{\mathbb{I}}_{\2}^{\left( 2\right) }$, and this suffices the role of the parity operator in the intrinsic entanglement analysis of Dirac spinors. Just to complement, the role of the parity operator could also be replaced by the results from the projection of the chiral operator. Of course, in both cases, the intrinsic entanglement would vanish for massless particle states.

To summarize, the correlations driven by the discrete spin-parity degrees of freedom of Dirac bispinors, as worked out in this paper, follow the $SU(2) \otimes SU(2)$ intrinsic symmetry properties as discussed above.

\section{Analytical Dirac cat states in a magnetic field}
\label{AppB}
The eigenfunction basis in Eqs.~(\ref{9998})-(\ref{9999}) was given simultaneously for both signs of the Hamiltonian for a charged fermion, corresponding to the negative and positive charges. The charge sign affects the center of the motion along the gauge-dependent coordinate, which is given by
\begin{equation}
s_r = \sqrt{e {\mathcal B}} \left( x + (-1)^r\frac{ k_y}{ e{\mathcal B}}
\right).
\end{equation}
To study the temporal evolution of any quantum superposition with both types of states, one needs first to fix the charge sign in the Hamiltonian. Assuming the Hamiltonian for a negative charge, here described by a positive intrinsic parity $r=1$, the negative energy solutions simply correspond to the previous negative intrinsic parity $r=2$, where one keeps the same basis notation for spin projection \cite{Greiner}. However, all eigenfunctions are now centered at the same $s_{r=1}$ coordinate, and $s_r \equiv s$ shall be adopted from now on. 

The time evolution of energy-localized quantum states are fully determined by the excited eigenstates. This follows from the eigenfunction expansion with coefficients determined by the non-stationary state $ \phi (s,t=0) $ as 
\begin{equation}\label{coeff}
 c_{n,r} ^\nu= \int ds \, \phi^\dagger (s,t=0) u _{n,r} ^\nu (s),
\end{equation}
where the integral over other spatial coordinates is trivial. Even though most overlap integrals may only be calculated efficiently using appropriate numerical methods, symmetries of the initial state determine the selection rule for the integral above. One considers, for instance, a state that can be written in the product form
\begin{equation}
 \phi (s,t=0) = f(s) \upphi,
\end{equation}
where it is assumed that $f(s)$ is a scalar function with well-defined parity and $ \upphi$ is a constant spinor.

Since the Hamiltonian eigenstates do not have a definite spin direction except for the lowest Landau level, they are in a superposition of eigenstates of the spin operator $\Sigma_z = diag\{ \sigma_z,\sigma_z\} $, each with a spatial parity symmetry. Therefore, if $ \upphi$ is an eigenstate of the spin operator, the only non-vanishing integrals in Eq.~\eqref{coeff} correspond to Hamiltonian eigenstates with quantum numbers $n$ of the same parity. Otherwise, these quantum numbers have no parity, even if $f(s)$ has spatial parity symmetry.

Dirac cat states proposed in \cite{PRA2021} satisfy the first case described above with analytical expansion coefficients, 

\begin{eqnarray}
c_{m+1,r=1} ^{+} &=& \mathcal{N}_a \exp(-a^2/4) (a/\sqrt{2}) ^{m} / \sqrt{m!} \label{levela0}, \\
c_{m+1,r=2} ^{+} &=& \mathcal{N}_a (-A_m) \exp(-a^2/4) (a/\sqrt{2}) ^{m} / \sqrt{m!} \label{levela1}, \\
c_{m+1,r=2} ^{-} &=& \mathcal{N}_a B_m \exp(-a^2/4) (a/\sqrt{2}) ^{m} / \sqrt{m!} \label{levela2},
\end{eqnarray}
where $\mathcal{N}_a$ is the normalization constant and $a$ the distance parameter.

If $m=2k$, for integer $k$, $\mathcal{N}_a = \cosh ^{-1/2} (a^2/2)$. The non-vanishing component of the Dirac spinor at $t=0$ for even Landau levels is given by $ \left(\begin{matrix} 1 & 0 & 0& 0 \end{matrix}\right)^T$ multiplied by
\begin{equation}\label{simp}
 \exp(-a^2/4) \,\sum_{n=0} ^{\infty} \mathcal{F}_ {2n} (s) \frac{(a/\sqrt{2})^{2n}}{\sqrt{(2n)!}} = \left( \frac{e \mathcal{B}}{\pi}\right)^{1/4} e^{-s^2/2} \sum_{n=0} ^{\infty} \frac{ H_{2n} (s) }{(2n)!} \left(\frac{a}{2}\right)^{2n},
\end{equation}
or more simply,
\begin{equation}\label{catstate}
\phi^S (s, t=0) = \frac{1}{2}\left( \frac{e \mathcal{B}}{\pi} \right)^{1/4} \bigg \{ \exp\left[-\frac{1}{2}(s-a)^2 \right] + \exp\left[-\frac{1}{2}(s+a)^2 \right] \bigg \} \left(\begin{matrix} 1 & 0 & 0& 0 \end{matrix}\right)^T,
\end{equation}
where $S$ stands for a symmetric cat state, which is even on the continuous variable $s$.

In contrast, if $m=2k+1$, for integer $k$, one has
\begin{equation}\label{catstatea}
 \phi^A (s, t=0) = \frac{1}{2}\left( \frac{e \mathcal{B}}{\pi} \right)^{1/4} \bigg \{ \exp\left[-\frac{1}{2}(s-a)^2 \right] - \exp\left[-\frac{1}{2}(s+a)^2 \right] \bigg \} \left(\begin{matrix} 1 & 0 & 0& 0 \end{matrix}\right)^T.
\end{equation}
where $A$ stands for an antisymmetric cat state, with $\mathcal{N}_a = \sinh ^{-1/2} (a^2/2)$. This state is odd on the continuous variable $s$.

The temporal evolution of $\phi^{A,S}(s,t)$ is fully encoded in the phase of each time-evolved eigenfunction. The parity of the principal quantum number in either case will determine an interesting collective behavior to be further explored. 

\section{Dynamical evolution of spinor matrix operators for ensembles of cat states}
\label{AppC}
    The computation of the mean values of quantum operators from Eq.~\eqref{V4} follows from the same arguments used for evaluating the localized probability density. For block diagonal constant matrices, the selection rule yields $n=m$, that is, the only nonvanishing contribution comes from states with the same HO quantum number. In contrast, for block off-diagonal matrices, averaged values always vanish for states with parity-defined principal quantum number. Thus, the $n=m$ components will be computed for both cases. The prefactor for even contributions,
\begin{equation}
\frac{4}{\cosh{ (a^2/2)}} \sum_{n} \frac{(a^2/2)^{2n}}{(2n)!},
\end{equation}
will be temporarily omitted. In a similar fashion, odd states are straightforwardly computed by summing over the odd components,
\begin{equation}
\frac{-4}{\sinh{ (a^2/2)}} \sum_{n}\frac{(a^2/2)^{2n+1}}{(2n+1)!}.
\end{equation}
It means that each component is summed over with the appropriate weighing factor for this particular ensemble. The spatial components of the Dirac current (cf. Eq.~\eqref{diraccurrent}) yield
\begin{eqnarray}
g_1 ^z &=& \frac{4 M \eta_n A_n}{E_n} \sin ^2(E_n t) \mathcal{F} ^2 _{n-1} (s), \\
g_1 ^y &=& 4 \eta_n ^2 B_n (1 + A_n ^2 + B_n ^2) \sin(E_n t)\cos(E_n t) \mathcal{F} _{n} (s) \mathcal{F} _{n-1} (s), \\ 
g_1 ^x &=& 4 \eta_n ^2 B_n (-1 + A_n ^2 + B_n ^2) \sin ^2(E_n t)\mathcal{F} _{n} (s) \mathcal{F} _{n-1} (s),
\end{eqnarray}
where the dependence on the variables $(s,t)$ is implied. These are simply the components of the velocity operator whose $x,y$ components are symmetric after averaging over the remaining variable. In a similar fashion, the time and spatial (Eq.~\eqref{spindensity}) components of the pseudo-vector density read
\begin{eqnarray} 
f_1 &=& - 4 \eta_n ^2 A_n ( -1 + A_n ^2 + B_n ^2) \sin ^2(E_n t) \mathcal{F} ^2 _{n-1} (s) \label{chiral}, \\
g_0 ^z &=& \mathcal{F} ^2 _{n-1} (s) -4\eta_n ^2 B_n ^2 \sin ^2(E_n t)(\mathcal{F} ^2 _{n-1} (s) + \mathcal{F} ^2 _{n} (s) ), \\
g_0 ^y &=& 0, \\ 
g_0 ^x &=& - 8 A_n B_n \eta_n ^2 \sin ^2(E_n t) \mathcal{F} _{n} (s) \mathcal{F} _{n-1} (s),
\end{eqnarray}
which describe the temporal evolution of the $\gamma_5$ and spin operators. Again, only the first and second lines correspond to even functions that do not vanish after integration. 

The scalar $f_3 (s,t)$ and pseudoscalar $f_2 (s,t)$ components are also described by even functions, 
\begin{eqnarray}
f_2 &=& -2 A_n \eta_n \sin(2 E_n t)\mathcal{F}_{n-1} ^2(s), \\
f_3 &=& 1- 8 \sin^2(E_n t) \eta_n ^2 (A_n^2 \mathcal{F}_{n-1} ^2(s) + B_n ^2 \mathcal{F}_{n} ^2 (s)) .
\end{eqnarray}

Finally, the antisymmetric tensor density is split into the components ${\bf g}_2 (s,t)$,
\begin{eqnarray}
g_2 ^z &=& - 2A_n \eta_n \sin(2E_n t) \mathcal{F}_{n-1} ^2 (s),\\
g_2 ^y &=& -4B_n (-1 + A_n^2 +B_n ^2) \eta_n ^2 \mathcal{F}_{n-1}(s)\mathcal{F}_{n} (s) \sin^2(E_n t) \\
g_2 ^x &=& 2B_n (1 + A_n^2 +B_n ^2) \eta_n ^2 \mathcal{F}_{n-1} (s)\mathcal{F}_{n} (s) \sin(2E_n t),
\end{eqnarray}
and the components ${\bf g}_3 (s,t)$, where
\begin{eqnarray}
g_3 ^z & =& \mathcal{F} ^2 _{n-1} (s) -4\eta_n ^2 \sin ^2(E_n t)\bigg(\mathcal{F} ^2 _{n-1} (s) (2 A_n ^2 + B_n ^2) - \mathcal{F} ^2 _{n} (s) B_n ^2 \bigg), \nonumber \\
g_3 ^x & =& 8 A_n B_n \sin^2(E_n t) \mathcal{F}_{n-1} (s)\mathcal{F}_{n} (s), \\
g_3 ^y & =& 0.
\end{eqnarray}

The $s-$integration of the functions above is immediate, since the basis $\mathcal{F}_n (s)$ is orthonormal. By recollecting the weighing factors, the only non-zero expectation values are thus
\footnotesize
\begin{eqnarray}
(1/4)\int \, ds \, g_1 ^z (s,t) &=& \langle \alpha_z \rangle _{CS} = \frac{4M}{\cosh{(a^2/2)}} \sum _n \frac{(a^2/2)^{2n}}{(2n)!}\frac{\eta_{2n+1} A_{2n+1}}{ E_{2n+1}} \sin({E_{2n+1}t})^2 ,\\
(1/4) \int \, ds \, g_0 ^z (s,t) &=& \langle \gamma_5 \alpha_z \rangle _{CS} = 1 - \frac{8}{\cosh{(a^2/2)}} \sum _n\frac{(a^2/2)^{2n}}{(2n)!} \eta_{2n+1} ^2 B_{2n+1}^2 \sin({E_{2n+1}t})^2 ,\\ 
(1/4) \int \, ds \, g_2 ^z (s,t) &=& - \langle i \gamma_z \rangle _{CS} = \frac{2}{\cosh{(a^2/2)}} \sum _n\frac{(a^2/2)^{2n}}{(2n)!} \eta_{2n+1} A_{2n+1} \sin(2 {E_{2n+1}t}), \\
(1/4) \int \, ds \, g_3 ^z (s,t) &=& - \langle \gamma_5 \gamma_z \rangle _{CS} = 1 - \frac{8}{\cosh{(a^2/2)}} \sum _n\frac{(a^2/2)^{2n}}{(2n)!} \eta_{2n+1} ^2 A_{2n+1}^2 \sin({E_{2n+1}t})^2, \\
(1/4) \int \, ds \, f_3 (s,t) &=& \langle \gamma_0 \rangle _{CS} = 1 - \frac{8}{\cosh{(a^2/2)}} \sum _n\frac{(a^2/2)^{2n}}{(2n)!} \eta_{2n+1} ^2 (A_{2n+1}^2 + B_{2n+1}^2) \sin({E_{2n+1}t})^2,
\end{eqnarray}
\normalsize
with $\langle i \gamma_z \rangle _{CS} = \langle i \gamma_0 \gamma_5 \rangle _{CS}$ and $\langle \gamma_5 \rangle _{CS} = - \langle \alpha_z \rangle _{CS}$, where $\langle \cdot \rangle_{CS}$ corresponds to the ensemble average for symmetric cat states.

\end{document}